\documentclass[]{pasj01} 
\Received{$\langle$20--Aug--2021$\rangle$}
\Accepted{$\langle$08--Nov--2021$\rangle$}
\Published{$\langle$publication date$\rangle$}

\usepackage{longtable}
\usepackage{graphicx}
\usepackage{txfonts}
\usepackage{natbib}
\usepackage{subfloat}
\usepackage{lineno}
\def\la{\ifmmode{\lesssim}\else$\lesssim$\fi}
\def\ga{\ifmmode{\gtrsim}\else$\gtrsim$\fi}

\begin{document}

\title{Star burst in W49N presumably induced by cloud-cloud collision}

\author{Ryosuke Miyawaki$^1$, Masahiko Hayashi$^{2,3}$, and Tetsuo Hasegawa$^2$}%
\altaffiltext{1}{College of Arts and Sciences, J.F. Oberlin University, Machida, Tokyo 194-0294, Japan}
\email{miyawaki@obirin.ac.jp}
\altaffiltext{2}{National Astronomical Observatory of Japan, 
2-21-1 Osawa, Mitaka, Tokyo 181-8588, Japan}
\altaffiltext{3}{JSPS Bonn Office, Ahrstr. 58, 53175 Bonn, Germany}


\maketitle

\begin{abstract}\label{Abstract}

We present high resolution observations of CS ($J=1-0$), H$^{13}$CO$^+$ ($J=1-0$), and SiO ($v=0: J=1-0$) lines, together with the 49~GHz and 86~GHz continuum emissions, toward W49N carried out with Nobeyama Millimeter Array.
We identified 11 CS, 8 H$^{13}$CO$^+$, and 6 SiO clumps with radii of 0.1--0.5~pc.
The CS and H$^{13}$CO$^+$ clumps are mainly divided into two velocity components, one at 4~km\,s$^{-1}$ and the other at 12~km\,s$^{-1}$, while the SiO clumps have velocities between the two components.
The SiO emission is distributed toward the UCHII ring, where the 4~km\,s$^{-1}$ component clumps of CS and H$^{13}$CO$^+$ also exist.
The 12~km\,s$^{-1}$ component clumps of CS are detected at the east and west of the UCHII ring with an apparent hole toward the ring.
The clump masses vary from 4.4$\times$10$^2$~M$_{\odot}$ to 4.9$\times$10$^4$~M$_{\odot}$ with the mean values of 
0.94$\times$10$^4$~M$_{\odot}$, 0.88$\times$10$^4$~M$_{\odot}$, and 2.2$\times$10$^4$~M$_{\odot}$ for the CS, H$^{13}$CO$^+$, and SiO clumps, respectively.
The total masses derived from CS, H$^{13}$CO$^+$, and SiO clumps are 1.0$\times$10$^5$~M$_{\odot}$, 0.70$\times$10$^5$~M$_{\odot}$, and 1.3$\times$10$^5$~M$_{\odot}$, respectively, which agree well with the corresponding virial masses of 0.71$\times$10$^5$~M$_{\odot}$, 1.3$\times$10$^5$~M$_{\odot}$, and 0.88$\times$10$^5$~M$_{\odot}$, respectively.
The average molecular hydrogen densities of the clumps are 0.90$\times$10$^6$\,cm$^{-3}$, 1.4$\times$10$^6\,$cm$^{-3}$, and 7.6$\times$10$^6\,$cm$^{-3}$ for the CS, H$^{13}$CO$^+$ and SiO clumps, respectively.
The density derived from the SiO clumps seems significantly higher than those from the others, probably because the SiO emission is produced in high density shocked regions.
The free fall time scale of the clumps is estimated to be $\sim$3$\times$10$^{4}$~yr, which gives an accretion rate of 3$\times$10$^{-3}$--1~M$_\odot\,$yr$^{-1}$ onto a stellar core.
The observed clumps are, if they are undergoing free fall, capable of producing dozens of massive stars in the next 10$^5$~yr.
We propose a view that pre-existing two clouds having the radial velocities of 4~km\,s$^{-1}$ and 12~km\,s$^{-1}$ collided with each other almost face-on to produce the observed clumps with intermediate velocities and triggered the burst of massive star formation in W49N.
\end{abstract}

\section{Introduction}
\label{Introduction}

The thermal radio source W49A is one of the most luminous HII region/molecular cloud complexes in our Galaxy located at a distance of 11.11$^{+0.79}_{-0.69}$ kpc \citep{Zhang2013}. 
Over an area of 15~pc ($\sim5\arcmin$) in diameter, it contains numerous compact and ultracompact HII (UCHII) regions, mainly grouped into W49 North (hereafter called W49N) and W49 South \citep[cf.][]{Dreher1984, DePree1997}. 
The HII regions as a whole emit $\sim$10$^{51}$ Lyman continuum photons per second, equivalent to the number of photons emitted by $\sim$100 O stars  \citep{Alves2003}.

High resolution continuum observations of W49N at centimeter and millimeter wavelengths have revealed clusters of ultra-compact HII regions (UCHII ring) located along an ellipse, or a tilted ring, with a diameter of 1.5~pc ($\sim25\arcsec$) \citep{Welch1987}.
The source~G, which in itself consists of at least several UCHII regions \citep[e.g.,][]{Dreher1984, DePree1997}, is the brightest among them.
It is also the most luminous H$_2$O maser source in the Galaxy \citep[e.g.,][]{Walker1982}. 

The molecular gas associated with W49A was first identified by \citet{Scoville1973} and was mapped in the CO ($J=1-0$) line at a resolution of 1$\arcmin$ \citep{Mufson1977}.
Recent observations of W49A in the $^{13}$CO ($J=1-0$) and C$^{18}$O ($J=1-0$) lines show 14 discernible features (MHH-1 to MHH-14), most of which are distributed within an area 20~pc ($\sim6\arcmin$) in diameter with their masses totaling 1.7$\,\times\,$10$^6$ M$_\odot$ \citep{Miyawaki2009}.
The feature MHH-1, corresponding to W49N, is unique in that it exhibits a large velocity width (15 km\,s$^{-1}$ FWHM) and is compact in size (2.3 pc $\times$ 3.0 pc FWHM), while containing a large mass of 2.4$\,\times\,$10$^5$ M$_\odot$. 
W49N also contains hot molecular cores characterized by emissions from molecules such as CH$_3$CN or SiO \citep[e.g.,][]{Wilner2001, Miyawaki2002}. 
The large mass and compact size of W49N are also confirmed by other molecular line observations \citep{Miyawaki1986, Serabyn1993}.

There are mainly two views, not exclusive to each other, to interpret these observations.
\citet{Welch1987} proposed from interferometric HCO$^+$ observations that a massive molecular envelope is freely infalling toward the UCHII ring at the center of W49N. 
The presence of infalling gas is consistent with the absorption features of H$_2$CO at 2 cm and 6 cm \citep{Dickel1990}.
\citet{Jackson1994} found,  from ${\rm NH_3}$ observations, warm molecular gas associated with the UCHII ring, suggesting that the ${\rm NH_3}$ velocity field 
reveals systematic radial motions at an infall velocity of $\sim$10 ~km\,s$^{-1}$.

 \citet{Miyawaki1986} proposed the other view that collision of two clouds at $V_{\rm LSR}\,\sim$4~km\,s$^{-1}$ and at $\sim$12 km\,s$^{-1}$ may have triggered a burst of star formation in W49N.
 \citet{Serabyn1993} identified, from multi-transition studies of CS, three dense clumps with a few $\times\,10^4\,$M$_\odot$ each in W49N, one at $\sim$4~km\,s$^{-1}$ at the center and the other two at $\sim$12~km\,s$^{-1}$ on its northeast and southwest sides, suggesting that cloud-cloud collision is a reasonable account.
Higher resolution CS ($J=2-1$) observations with the BIMA array basically follow the same tendency for the two velocity components \citep{Dickel1999}.
The interferometric CS  ($J=2-1$) spectra have redshifted absorption features toward source~G, indicative of infall motions, although the infalling interpretation may not be unique (Williams et al. 2004). 

Although there is some evidence to support the global collapse model of W49N \citep[][]{Welch1987, Williams2004}, the cause of such a sytematic infall, and the very high star formation activities in W49N as well, has not yet been well understood. 
It is also conceivable that the massive clouds of $\sim10^5$~M$_\odot$ each (MHH-1 to MHH-14) distributed in a larger W49A area have a reasonable chance to collide with each other \citep{Miyawaki2009}, when we consider their mean free time of 10$^6$~yr per cloud.
A gravitationally unstable core may have formed through such collisions of massive clouds, leading to the free fall to form the clusters of UCHII regions with $\sim$100 O stars.
Higher resolution observations of gas around the UCHII ring are necessary to examine such a hypothesis.

In this paper we present Nobeyama Millimeter Array (NMA) observations of the massive molecular cloud core W49N with the transitions of CS ({{\it J}\,=\,1\,--\,0}), SiO ({{\it J}\,=\,2\,--\,1}), and H$^{13}$CO$^+$ ({{\it J}\,=\,1\,--\,0}) together with the 49~GHz and 86~GHz continuum emissions.
We show high resolution images and discuss star formation in the massive core of W49N in relation to the UCHIIs and hot molecular cores, providing further evidence of cloud-cloud interaction.
The nomenclature in this paper, except for newly found sources, is taken from the previous studies by \citet{Dreher1984}, \citet{Dickel1990}, and \citet{Wilner2001}.

\section{Observations}\label{Observations}

\subsection{CS emission}\label{Obs-CS}

The CS  ($J=1-0$) observations were made from April 1988 to April 1989 on B, C and D configurations of the Nobeyama Millimeter Array (NMA),  consisting of six 10-m antennas, at NRO.\footnote{Nobeyama Radio Observatory, a branch of the National Astronomical Observatory, National Institutes of Natural Sciences}
The field center was at (${\alpha(1950)}$,  ${\delta(1950)}$) = (19$^{\rm h}\,$07$^{\rm m}\,$49\,\fs8, 9\degree\,01$'$\,17\farcs1) with a primary beam size of $\sim150''$ (FWHM) at 49 GHz.
The receivers were equipped with cryogenically cooled  Superconductor-Insulator-Superconductor (SIS) mixers with a system noise temperature of $\sim$200 K in double side band operation, tuned to the frequency of CS ($J=1-0$) (48.990967 GHz).
A 1024 channel digital FFT spectro-correlator (FX) covered a 80 MHz bandwidth with the velocity resolution of 0.48~km\,s$^{-1}$ corresponding to the channel width of 78 kHz.
Bandpass and flux calibrations were carried out in the usual manner with 3C84 and 1749+096, respectively.
The minimum and maximum baseline lengths were 20 m (3.1 k$\lambda$) and 210 m (32 k$\lambda$), respectively.
We applied natural weighting, and the resultant synthesized beam size was 9\farcs2$\times$ 6\farcs9 (PA=176\degree.4).
The RMS noise level of each channel map is 0.05\,Jy\,beam$^{-1}$, which corresponds to 0.40\,K in brightness temperature.
Most of the emission features are located within $\sim30''$ from the map center.
No correction for the primary beam attenuation was applied.

\subsection{SiO and H$^{13}$CO$^+$ emissions}
\label{Obs-SiO and H13CO+}

Observations of SiO ($J=2-1$) and H$^{13}$CO$^+$ ($J=1-0$) were made from January to April 1991 on C and D configurations of NMA.
The field center was at (${\alpha(1950)}$, ${\delta(1950)}$) = (19$^{\rm h}\,$07$^{\rm m}\,$49\,\fs8,  9\degree\,01$'\,$15\,\farcs5) with a primary beam diameter of $\sim80''$ (FWHM) at 86 GHz.
The antennas were equipped with SIS receivers having system noise temperature of $\sim$300 K.
The FX had a simultaneous bandwidth of 320~MHz, covering both SiO (v=0: $J=2-1$) (86.846998 GHz) and H$^{13}$CO$^+$ ($J=1-0$) (86.754330 GHz) lines at the same time with the velocity resolution of 1.08~km\,s$^{-1}$ corresponding to the channel width of 312 kHz.
Bandpass  and flux calibrations were carried out in the usual manner with 3C273 and 1749+096, respectively.
The minimum and maximum baseline lengths were 20 m (5.7 k$\lambda$) and 135 m (38 k$\lambda$), respectively.
We applied uniform weighting, and the resultant synthesized beam size was 4\farcs7$\times\,$3\farcs7 (PA=$-$65\degree.2).
The RMS noise level in each channel map is 0.05\,Jy\,beam$^{-1}$, which corresponds to 0.46\,K in brightness temperature.
The farthest features are located $\sim20''$ away from the map center.
No correction for the primary beam attenuation was applied.

\subsection{Continuum emission}\label{Obs-Cont}

We measured the visibilities of the continuum emission by integrating line free channels of the FX data.
The continuum visibilities thus determined were subtracted from all the FX channels to obtain baselines of line spectra.
We used the NRAO AIPS package to produce the maps.
Uniform weighting was applied for the continuum emission, and the resultant synthesized beam sizes were 5\farcs6$\times$ 4\farcs3 (PA=$-$8\degree.0) and 4\farcs7$\times$ 3\farcs7 (PA=$-$65\degree.2) for the 49 GHz and 86 GHz continuum emissions, respectively.

The beam sizes and noise levels are listed in Table~\ref{Table01} together with molecular parameters used later (see \S4.2).
All the maps were converted from the B1950 to J2000 coordinates.
We then processed the data using the standard packages of CASA  \citep{McMullin2007}.

\begin{table*}[ht]

\begin{minipage}{\textwidth}
\caption{Observed molecular lines and continuum}
\label{Table01}
\begin{center}
\scalebox{0.85}[0.85] 
{

\begin{tabular}{llccccccccc}
\hline\hline

Molecule  & Transition  & $\nu_0$ & $E_\mathrm{u}$  & $\mu$       & $A_{ul}$   &    $B$   & Abundance  &  $\theta_\mathrm{beam}$ &PA              & RMS noise\\

                &                   & (GHz)    & (K)                        &   (Debey)  & (s$^{-1}$) &(GHZ)  &  &($''$)                                 &   (\arcdeg)  & (Jy\,beam$^{-1}$)\\
\hline
CS           & $J =1 - 0$          & 48.990954 & 2.3512  & 1.957&1.8 $\times$ 10$^{-6}$&22.45 & 2.2 $\times$ 10$^{-9}$ $^{a)}$& 9.2 $\times$ 6.9&176.4 & 0.05\\
H$^{13}$CO$^{+}$ & $J = 1 - 0$ & 86.754288 & 4.1635 &  3.88 & 3.9 $\times$ 10$^{-5}$&46.38 & 2.9 $\times$ 10$^{-11}$ $^{a)}$ &   4.7 $\times$ 3.7 & $-$65.2    &0.05\\
SiO          & $J = 2 - 1$   & 86.846985 & 6.2520 & 3.098& 2.9 $\times$ 10$^{-5}$&21.71 & 3.8 $\times$ 10$^{-11}$$^{b)}$ &   4.7 $\times$ 3.7 & $-$65.2 & 0.05\\
\hline
49~GHz continuum&                   &                   &             &           &                       &              &        &   5.6 $\times$ 4.3 &$-$8.0        &0.02\\
86~GHz continuum&                   &                    &            &           &                         &            &        &   4.7 $\times$ 3.7 &$-$65.2    &0.02\\
\hline
\vspace{3mm}
\end{tabular}
}
\footnotetext{$^{a)}$  \citet{Rodriguez-Baras2021}, $^{b)}$ \citet{Li2019}}
\\
\end{center}
\end{minipage}

\end{table*}


\section{Results}\label{Results}

\subsection{Continuum emission}\label{Result-Cont}

\begin{figure*}[htbp]
\includegraphics*[bb= 0 100 700 850, scale=0.75]{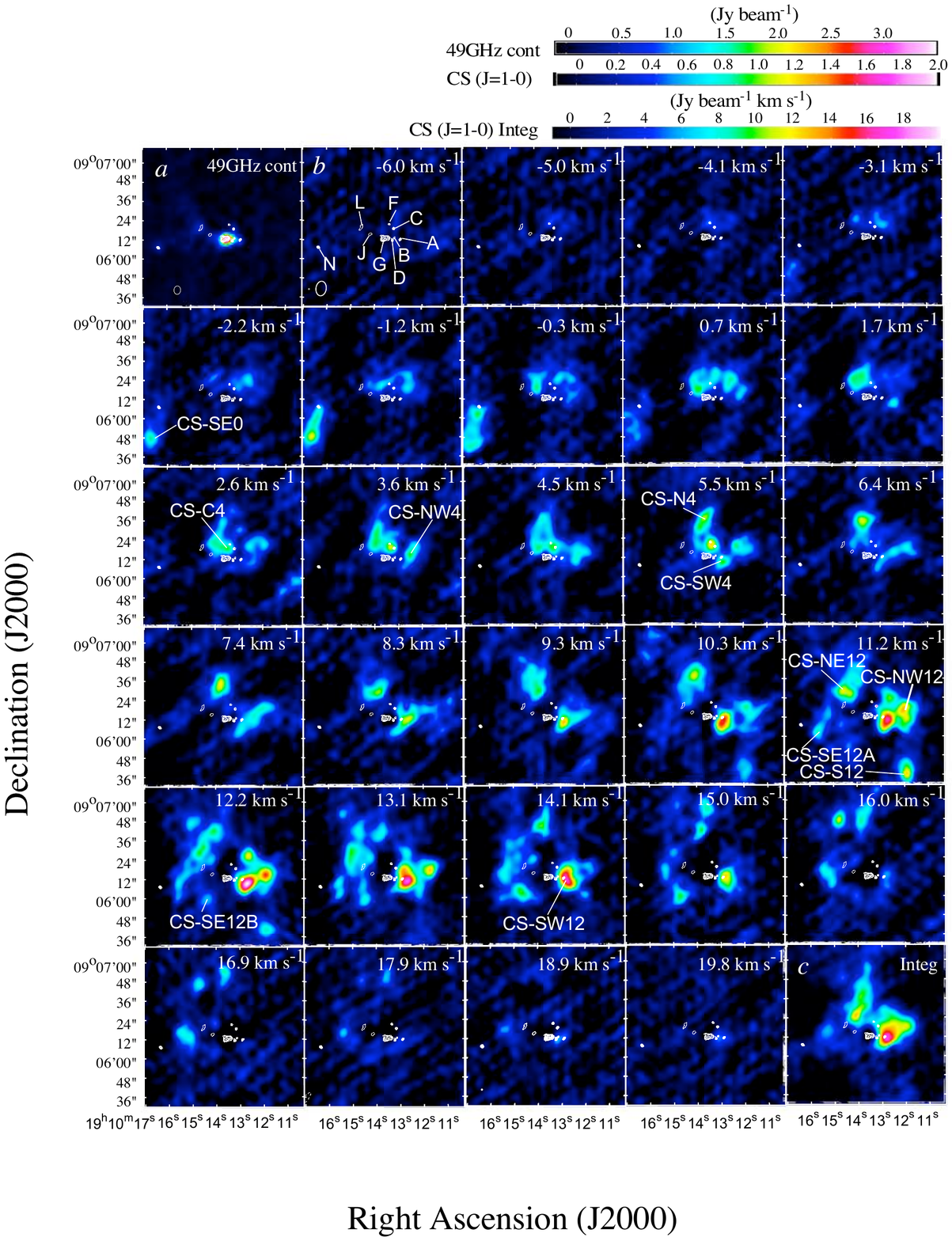}
\caption{
The 49~GHz continuum (top left panel $a$) and CS ($J=1-0$) (other panels) maps of W49N.
The integrated intensity map of CS is shown in the bottom right panel~$c$.
The other CS maps show the intensity averaged over two channels (0.96~km\,s$^{-1}$) around the LSR velocity noted at the upper right corner of each panel.
The beam sizes for the 49 GHz continuum and line observations are shown in panels $a$ and $b$, respectively.
The 8.3~GHz continuum image \citep[][]{DePree1997} is superposed on each panel in white contours at 20\%, 30\%, 40\%, 50\%, 60\%, 70\%, 80\%, and 90\% levels.
\label{Fig01}}
\end{figure*}

\begin{figure*}[htbp]
\includegraphics*[bb= 0 100 700 800, scale=0.75]{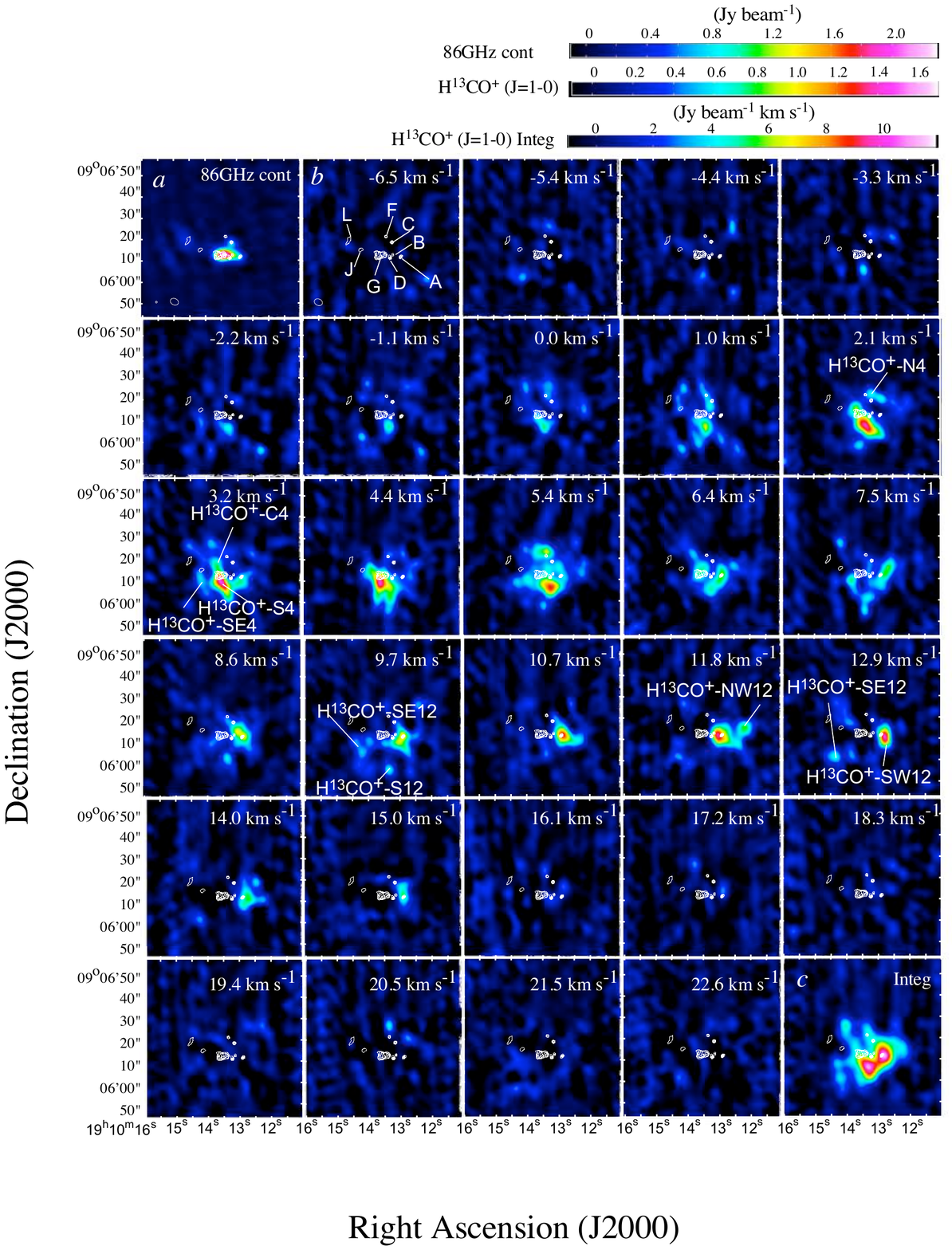}
\caption{
The 86~GHz continuum (top left panel $a$) and H$^{13}$CO$^+$ ($J=1-0$) (other panels) maps of W49N.
The integrated intensity map is shown in the bottom right panel~$c$.
The other H$^{13}$CO$^+$ maps show the intensity in the 1.08~km\,s$^{-1}$ channel width at the LSR velocity noted at the upper right corner of each panel.
The beam size is shown in panel $a$.
The 8.3~GHz continuum image \citep[][]{DePree1997} is superposed on each panel in white contours at 20\%, 30\%, 40\%, 50\%, 60\%, 70\%, 80\%, and 90\% levels.
\label{Fig02}}
\end{figure*}

\begin{figure*}[htbp]
\includegraphics*[bb= 0 100 700 800, scale=0.75]{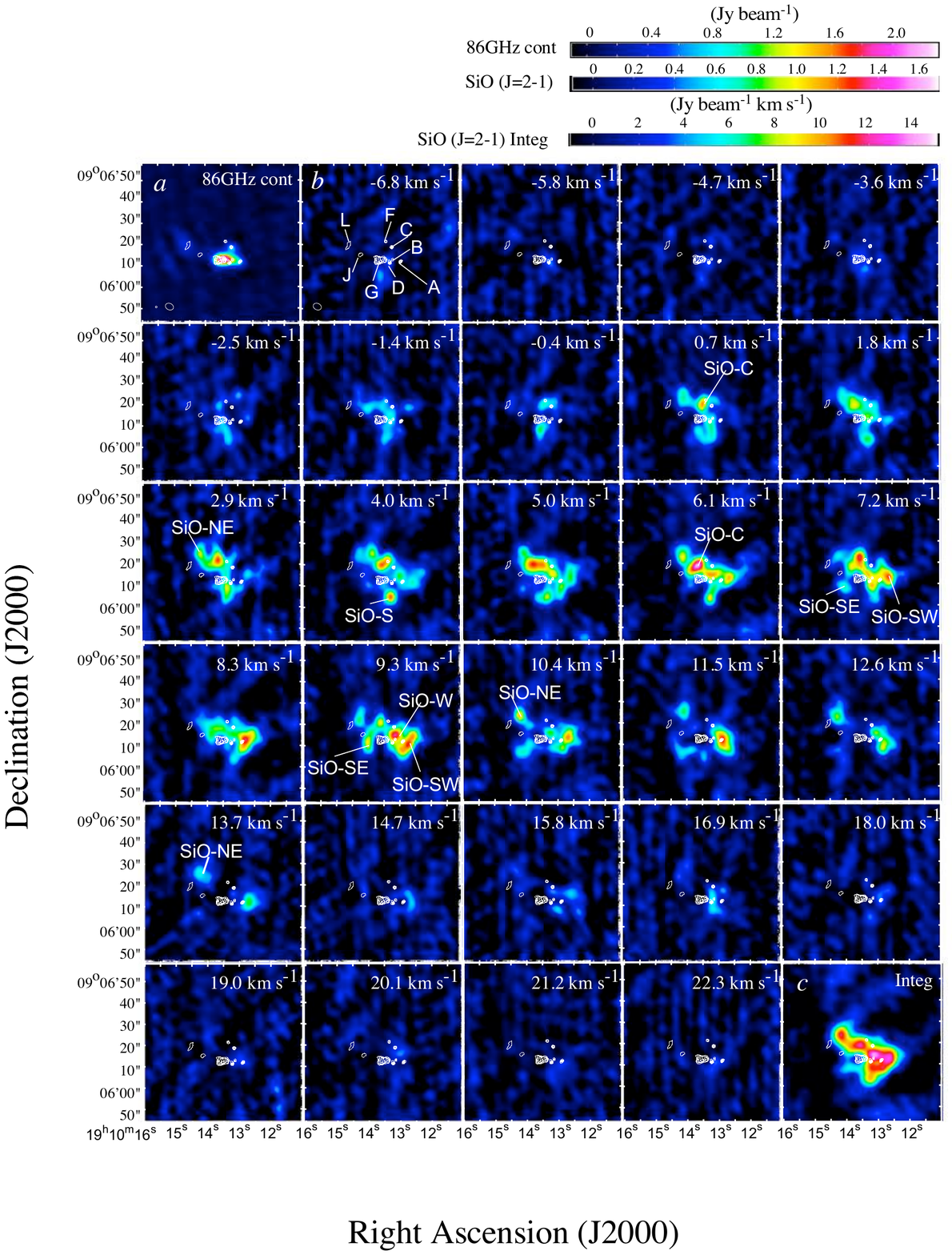}
\caption{
Same as Figure~\ref{Fig02}, but for SiO ($v=0:~J=2-1$).\label{Fig03}}
\end{figure*}

Figure~\ref{Fig01}$a$ (upper left panel) shows the 49 GHz continuum map, and Figures~\ref{Fig02}$a$ and \ref{Fig03}$a$ (upper left panels) both show the same 86 GHz continuum map.
The peak specific intensities are 3.4~Jy\,beam$^{-1}$ and 2.3~Jy\,beam$^{-1}$ for 49~GHz and 86~GHz, respectively.
The 8.3 GHz continuum map \citep[][]{DePree1997}, representing the UCHII regions A to G, J, L and N as noted in Figure~\ref{Fig01}$b$ \citep{Dreher1984}, are superposed in white contours on all the maps of Figures~\ref{Fig01}, \ref{Fig02} and \ref{Fig03}.

Both 49~GHz and 86~GHz continuum emissions are strongest at source~G with an extended feature to the west, i.e., to sources B and A, which, having rising spectra toward higher frequencies \citep{Wilner2001}, are not resolved well and do not have independent peaks.
The sources C, J, L and N have their weak counterparts at 49 GHz, while only C and L are seen above the noise level at 86 GHz.

\subsection{CS emission}\label{Result-CS}

Figure~\ref{Fig01} also shows the velocity channel maps of CS ($J=1-0$)  toward W49N, with an integrated intensity map in the bottom right panel $c$.
The peak specific intensity is 2.0~Jy\,beam$^{-1}$ occurring at $V_{\rm LSR}=$12.2~km\,s$^{-1}$.
As discussed in previous papers and Appendix~\ref{line profiles}, there are mainly two velocity components toward W49N at $\sim$4~km\,s$^{-1}$ and $\sim$12~km\,s$^{-1}$.
We call them as the 4~km\,s$^{-1}$ and 12~km\,s$^{-1}$ components, respectively, in this paper.

Examining the channel maps from low to high velocities, we recognize that the CS emission is distributed roughly in the UCHII ring at lower velocities ($-$1.2~km\,s$^{-1}\leq$  $V_{\rm LSR}\leq$~5.5~km\,s$^{-1}$), while the emission occurs mainly at two locations, one at the southwest of the ring near source~A and the other at the west of the ring, at higher velocities (6.4~km\,s$^{-1}\leq$  $V_{\rm LSR}\leq$~15.0~km\,s$^{-1}$).
This tendency is consistent with the previous results of CS and C$^{34}$S observations \citep{Serabyn1993}, but the NMA maps show by far more complicated distribution of CS emission in and around the UCHII ring.
Let us examine its distribution in more detail.

We may attribute four ``clumps'' to the 4~km\,s$^{-1}$ component.
They are identified in Figure~\ref{Fig01} as CS-C4, CS-N4, CS-NW4, and CS-SW4 by their positions and the suffix 4.
The clump CS-C4 is located inside the UCHII ring.
It corresponds to CS-C identified by \citet{Serabyn1993}.
CS-N4 is a feature located at the north of CS-C4 or UCHII ring and is seen at 2.6~km\,s$^{-1}\leq$ $V_{\rm LSR} \le$7.4~km\,s$^{-1}$.
Two other clumps are seen at the west and south of source~A.
CS-NW4 is mainly visible at 3.6~km\,s$^{-1}\leq$ $V_{\rm LSR}$ $\leq$5.5 ~km\,s$^{-1}$, and CS-SW4 is at 5.5~km\,s$^{-1}$.

We also identified six clumps for the 12~km\,s$^{-1}$ component in Figure~\ref{Fig01}: CS-NE12, CS-SE12A, CS-SE12B, CS-NW12, CS-SW12, and CS-S12.
CS-SW12, close to source~A, is the strongest CS emitting clump seen at 8.3~km\,s$^{-1}$ $\leq $$V_{\rm LSR}\leq$ 16.0 ~km\,s$^{-1}$.
It looks smoothly connected to CS-SW4 in velocity at its blueshifted side.
CS-NW12, located at the west of CS-SW12, is a major substructure of CS-SW identified by \citet{Serabyn1993}.
On the eastern side of the UCHII ring, there is CS-NE12 corresponding to CS-NE of \citet{Serabyn1993}.
The emission on this side is extended from north to south, surrounding the UCHII ring to overall form a C-shaped feature (see panels for 14.1~km\,s$^{-1}$ and 15.0~km\,s$^{-1}$) with CS-SE12A and CS-SE12B.
These two southern clumps were not detected by \citet{Serabyn1993}.

CS-S12, located far south, is seen at 10.3~km\,s$^{-1}$ $\leq$ $V_{\rm LSR}$ $\leq$12.2 ~km\,s$^{-1}$.
There is another clump, CS-SE0, which belongs to neither 4~km\,s$^{-1}$ nor 12~km\,s$^{-1}$ component seen at $-$3.1~km\,s$^{-1}$ $\leq$$V_{\rm LSR}$ $\leq$0.7 ~km\,s$^{-1}$.

\begin{figure}[htbp]
\includegraphics*[bb= 120 180 600 750, scale=0.55]{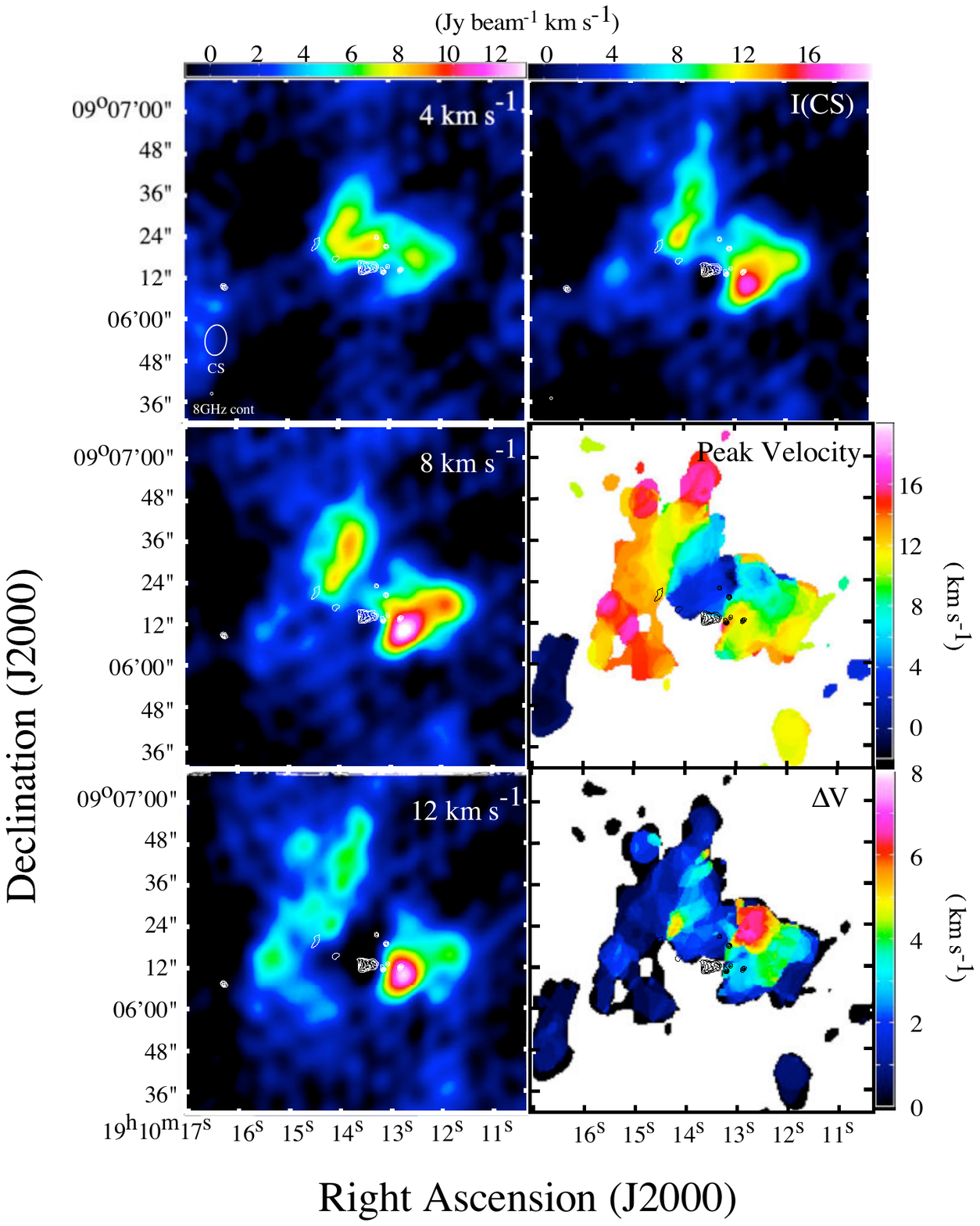}
\caption{
{\it Upper left:}
CS ($J=1-0$) intensity integrated from $V_{\rm LSR}=-$5.0~km\,s$^{-1}$ to 7.4~km\,s$^{-1}$.
{\it Center left:}
CS intensity integrated from 5.5~km\,s$^{-1}$ to 10.3~km\,s$^{-1}$.
{\it Lower left:}
CS intensity integrated from 8.3~km\,s$^{-1}$ to 19.8~km\,s$^{-1}$.
{\it Upper right:}
Total integrated intensity map CS emission (same as Figure~\ref{Fig01}$c$).
{\it Center right:} 
Peak velocity map.
{\it Bottom right:} 
Velocity width map.
The 8.3~GHz continuum image \citep[][]{DePree1997} is superposed in each panel in contours at 20\%, 30\%, 40\%, 50\%, 60\%, 70\%, 80\%, and 90\% levels.
Beam sizes are shown at the lower left corner of the top left panel.

\label{Fig04}}
\end{figure}

A summary of CS ($J=1-0$) results is presented in Figure~\ref{Fig04}: the CS intensity maps around 4~km\,s$^{-1}$, 8~km\,s$^{-1}$, and 12~km\,s$^{-1}$, and total integrated intensity (same as Figure~\ref{Fig01}$c$), peak velocity and velocity width maps.
These maps confirm that the emission at 4~km\,s$^{-1}$ overlaps with the UCHII ring, while the emission at 8~km\,s$^{-1}$ and 12~km\,s$^{-1}$ mainly arises from the northeast and southwest of the UCHII ring.
This is fully consistent with the result by \citet{Serabyn1993} that there are three clumps with one at 4~km\,s$^{-1}$ in the middle and the other two at 12~km\,s$^{-1}$ on its both sides.
In the peak velocity map, there is a clear velocity variation from the northeast to southwest along the main CS ridge that overlapped with the UCHII ring: namely, redshifted at the northeast, blueshifted at the center, and again redshifted at the southwest.
It cannot be regarded as a systematic motion, such as rotation, of a connected body.
The velocity width is wide at the north of source A, where no conspicuous feature is seen in the channel maps.

\subsection{ H$^{13}$CO$^+$ emission}\label{Result-H13CO+}

Figure~\ref{Fig02} shows the velocity channel maps of H$^{13}$CO$^+$($J=1-0$) emission, with an integrated intensity map in the bottom right panel $c$.
The peak specific intensity is 1.5~Jy\,beam$^{-1}$ occurring at 4.4 km\,s$^{-1}$.

We named 8 clumps by their positions and velocities: H$^{13}$CO$^+$-S4, H$^{13}$CO$^+$-C4, H$^{13}$CO$^+$-N4, and H$^{13}$CO$^+$-SE4 for the 4~km\,s$^{-1}$ component and H$^{13}$CO$^+$-SW12, H$^{13}$CO$^+$-NW12, H$^{13}$CO$^+$-SE12, H$^{13}$CO$^+$-S12 for the 12~km\,s$^{-1}$ component.

H$^{13}$CO$^+$-S4 is the most conspicuous clump at $-$3.3~km\,s$^{-1}$ $\leq$ $V_{\rm LSR}$ $\leq$5.4 ~km\,s$^{-1}$.
The three other weaker clumps for the 4~km\,s$^{-1}$ component are H$^{13}$CO$^+$-C4 at 3.2~km\,s$^{-1}$, corresponding to CS-C4, H$^{13}$CO$^+$-N4 at 2.1~km\,s$^{-1}$ and 5.4 ~km\,s$^{-1}$ located at the north of source C and F, and H$^{13}$CO$^+$-SE4 at 3.2~km\,s$^{-1}$ located at the south of source J .
For the 12~km\,s$^{-1}$ component, H$^{13}$CO$^+$-SW12 is the strongest clump located toward source A.
This clump is coincident with CS-SW12 and detected over a large velocity range at 6.4~km\,s$^{-1}$ $\leq$ $V_{\rm LSR}$ $\leq$15.0 ~km\,s$^{-1}$.
The other weaker 12~km\,s$^{-1}$ component clumps are H$^{13}$CO$^+$-S12 seen at  6.4~km\,s$^{-1}$ $\leq$ $V_{\rm LSR}$ $\leq$9.7~km\,s$^{-1}$, H$^{13}$CO$^+$-NW12 at 11.8~km\,s$^{-1}$, and H$^{13}$CO$^+$-SE12 at 9.7~km\,s$^{-1}$.

\begin{figure}[htbp]
\includegraphics*[bb= 120 180 600 750, scale=0.55]{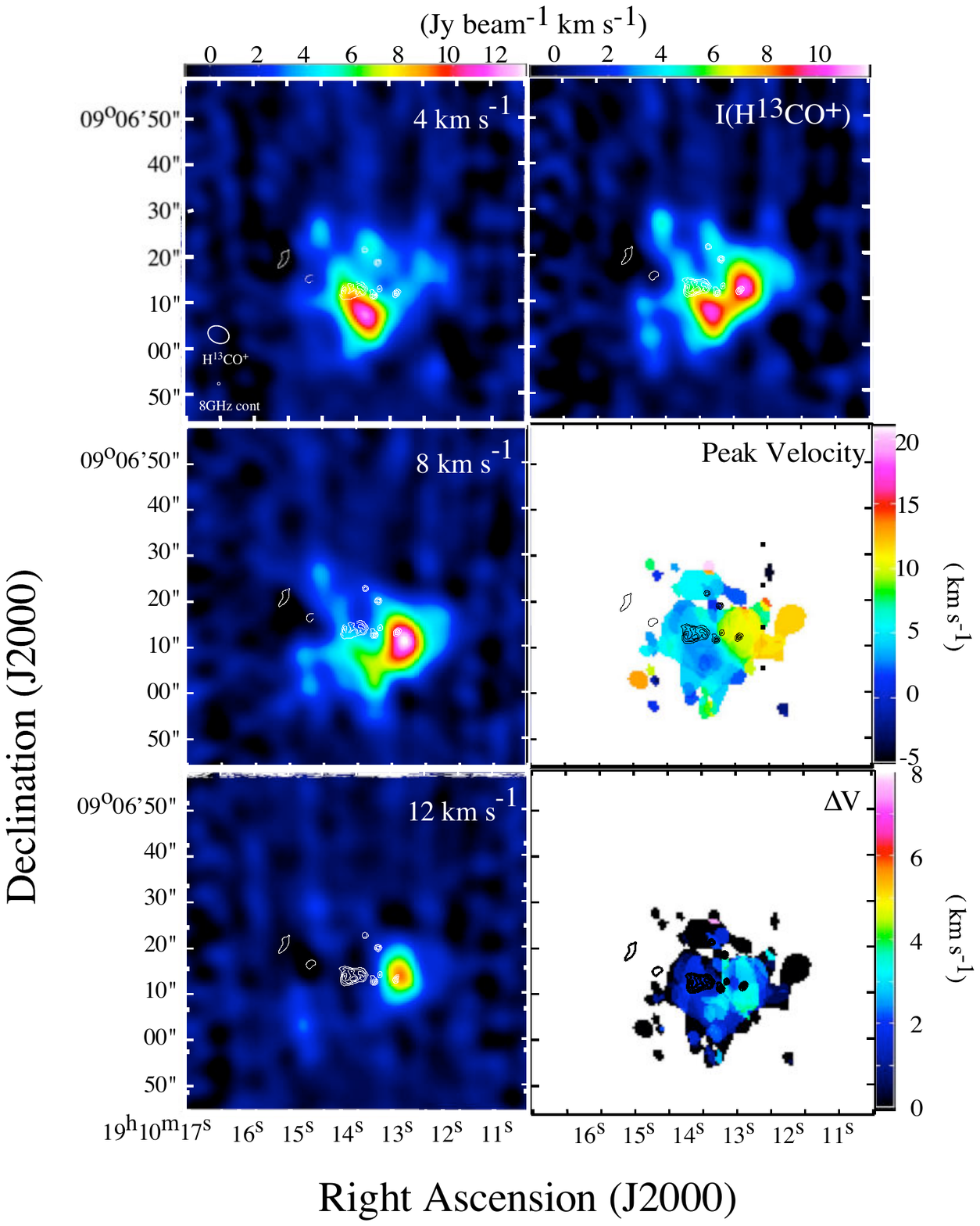}
\caption{
{\it Upper left:}
H$^{13}$CO$^+$ ($J=1-0$) intensity integrated from $V_{\rm LSR} = -$4.4~km\,s$^{-1}$ to 7.5~km\,s$^{-1}$.
{\it Center left:} 
H$^{13}$CO$^+$ intensity integrated from 5.4~km\,s$^{-1}$ to 10.8~km\,s$^{-1}$.
{\it Lower left:}
H$^{13}$CO$^+$ intensity integrated from 8.6~km\,s$^{-1}$ to 20.5~km\,s$^{-1}$.
{\it Upper right:} 
Total integrated intensity map of H$^{13}$CO$^+$ emission (same as Figure~\ref{Fig02}$c$).
{\it Center right:} 
Peak velocity map.
{\it Bottom right:} 
Velocity width map.
The angular resolution of 4\farcs7$\,\times\,$1\farcs85  (PA=-65\fdg 2).
The 8.3~GHz continuum image \citep[][]{DePree1997} is superposed in each panel in contours at 20\%, 30\%, 40\%, 50\%, 60\%, 70\%, 80\%, and 90\% levels.
Beam sizes are shown at the lower left corner of the top left panel.

\label{Fig05}}
\end{figure}

Figure~\ref{Fig05} summarizes the H$^{13}$CO$^+$ ($J=1-0$) results: the H$^{13}$CO$^+$ intensity maps around 4~km\,s$^{-1}$, 8~km\,s$^{-1}$, and 12~km\,s$^{-1}$, and total integrated intensity (same as Figure~\ref{Fig02}$c$), peak velocity and velocity width maps.
The H$^{13}$CO$^+$ emission mainly arises from the two strong clumps H$^{13}$CO$^+$-S4 and H$^{13}$CO$^+$-SW12 located south and southwest, respectively, of the UCHII ring.
In the peak velocity map, there is a clear trend of velocity gradient from the northwest to southeast along the ridge of H$^{13}$CO$^+$ emission at the southwest edge of the UCHII ring.
It is difficult to attribute the gradient to a rotating motion of a single gaseous body, because the gradient arises as a result of the two spatially separate clumps with different velocities (H$^{13}$CO$^+$-S4 and H$^{13}$CO$^+$-SW12) and there is not much emission connecting the two clumps at the intermediate velocities.

Comparing the spatial distributions of CS and H$^{13}$CO$^+$ emissions, we notice that there is an area that the two emissions are not spatially well correlated, especially for their 4~km\,s$^{-1}$ component at the south of source~G where only H$^{13}$CO$^+$ emission is seen.
The other 4~km\,s$^{-1}$ component clumps of CS (CS-N4, CS-C4 and CS-SW4) are, on the other hand, relatively well correlated with the H$^{13}$CO$^+$ emission at the north and west of the UCHII ring.
For the 12~km\,s$^{-1}$ component, the H$^{13}$CO$^+$ emission arises from the region where CS emission is strong: H$^{13}$CO$^+$-SW12 and CS-SW12 coincide with each other.

The difference in spatial distribution between the two molecular species may partly be due to the difference in their critical densities: $n_{\rm cr}$(H$_2$)$\,\sim\,$1.5$\times$10$^4$~cm$^{-3}$ for CS and $n_{\rm cr}$(H$_2$)$\,\sim\,$6.2$\times$10$^4$~cm$^{-3}$ for H$^{13}$CO$^+$ \citep{Shirley2015}.
It is not straightforward, however, to understand the absence of CS emission at the south of source~G, where strong H$^{13}$CO$^+$ emission (H$^{13}$CO$^+$-S4) is detected.
Such difference is also observed in other star forming regions \citep[e.g.,][]{Monge2014}.

\subsection{SiO emission}\label{Result-SiO}

Figure~\ref{Fig03} shows the velocity channel maps of SiO ($v=0:~J=2-1$) emission together with an integrated intensity map in the bottom right panel $c$.
The peak specific intensity is 1.6~Jy\,beam$^{-1}$ occurring at 6.1~km\,s$^{-1}$.

The integrated intensity map (Figure~\ref{Fig03}$c$) shows four main clumpy features aligned along the major axis of the UCHII ring.
We named the four clumps as SiO-NE, SiO-C, SiO-W and SiO-SW from northeast to southwest in the channel maps.
The clump SiO-C is located in the UCHII ring coinciding with CS-C4.
It covers the velocity range of $-$1.4~km\,s$^{-1}$ $\leq$ $V_{\rm LSR}$ $\leq$9.3~km\,s$^{-1}$.
SiO-NE positionally coincides with CS-NE12, but covers a larger velocity range from 0.7~km\,s$^{-1}$ to 13.7~km\,s$^{-1}$.
SiO-W is a clump located at the north of source~B prominent at 9.3~km\,s$^{-1}$ and is seen from 5.0~km\,s$^{-1}$ to 12.6~km\,s$^{-1}$.
SiO-SW is an elongated clump located at the west to the south of source~A.
It may be further divided into smaller clumps, but we call the elongated feature as a whole as SiO-SW for simplicity.
It positionally coincides with CS-SW12, but covers a larger velocity range from 1.8~km\,s$^{-1}$ to 13.7~km\,s$^{-1}$.

There are two other clumps identified on the velocity channel maps: SiO-S and SiO-SE.
SiO-S covers the velocity range of $-$3.6~km\,s$^{-1}$ $\leq$ $V_{\rm LSR}$ $\leq$8.3~km\,s$^{-1}$.
It appears as an extended feature from SiO-SW in the integrated intensity map (Figure~\ref{Fig03}$c$).
SiO-SE is seen at the southeast of source~G at 4.0~km\,s$^{-1}$ $\leq$ $V_{\rm LSR}$ $\leq$11.5~km\,s$^{-1}$.
It can be seen as a faint feature at the southeast of source~G in the integrated intensity (Figure~\ref{Fig03}$c$).

Examining each velocity channel map in detail, we notice that the distribution of SiO emission has a tendency to be spatially anti-correlated with the UCHII regions.
In the panel for $V_{\rm LSR}$ = 6.1~km\,s$^{-1}$, for example, the SiO clumps NE, C and SW align from northeast to southwest with the low level emission avoiding  each UCHII region.
As a result, the UCHII regions appear to be located at the waists of elongated wiggly SiO emission.
This is natural when observations with sufficiently high angular resolutions can spatially separate highly ionized gas from dense molecular gas.

\begin{figure}[htbp]
\includegraphics*[bb= 120 180 600 750, scale=0.55]{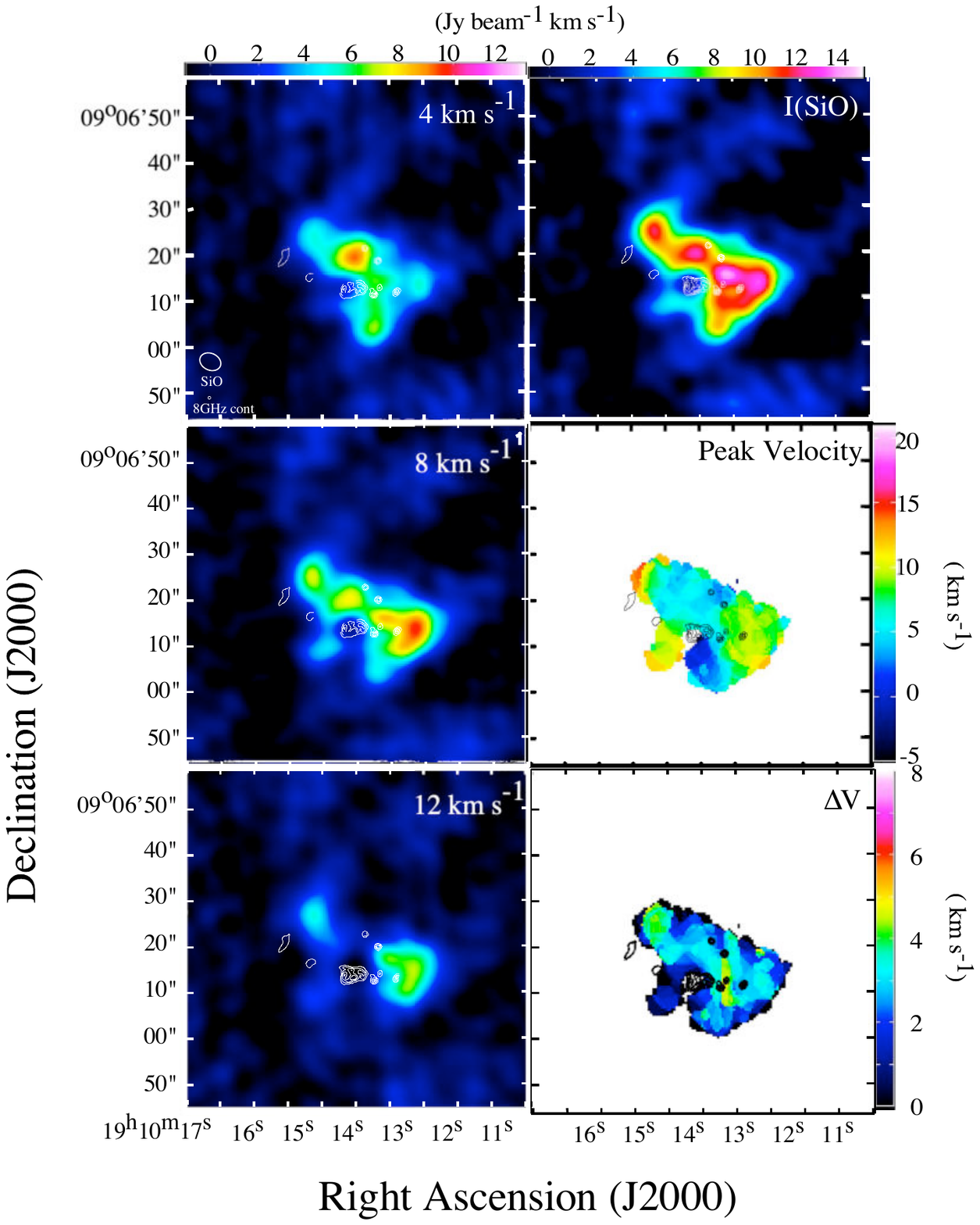}
\caption{
{\it Upper left:}
SiO ($v=0: J=2-1$) intensity integrated from $V_{\rm LSR} =-$4.7~km\,s$^{-1}$ to 7.1~km\,s$^{-1}$.
{\it Center left:} 
SiO intensity integrated from 5.0~km\,s$^{-1}$ to 10.4~km\,s$^{-1}$.
{\it Lower left:}
SiO intensity integrated from 8.3~km\,s$^{-1}$ to 20.1~km\,s$^{-1}$.
{\it Upper right:} 
Total integrated intensity map of SiO emission (same as Figure~\ref{Fig03}$c$).
{\it Center right:} 
Peak velocity map.
{\it Bottom right:} 
Velocity width map.
The angular resolution of 4\farcs7$\,\times\,$3.7\farcs85  (PA=-65\fdg 2).
The 8.3~GHz continuum image \citep[][]{DePree1997} is superposed in each panel in contours at 20\%, 30\%, 40\%, 50\%, 60\%, 70\%, 80\%, and 90\% levels.
Beam sizes are shown at the lower left corner of the top left panel.
\label{Fig06}}
\end{figure}

Figure~\ref{Fig06} summarizes the results of SiO ($v=0:~J=1-0$) emission: the SiO intensity maps around 4~km\,s$^{-1}$, 8~km\,s$^{-1}$, and 12~km\,s$^{-1}$, and integrated intensity (same as ~\ref{Fig03}$c$), peak velocity and velocity width maps.
A hole of SiO emission toward source~G is clear in all the intensity (moment 0) maps.
It is also apparent that the SiO emission directly toward source A tends to be weaker, showing a shallow dip, than its surrounding region.
These confirm the tendency discussed above that the dense molecular gas is separated from the highly ionized gas in the present high angular resolution NMA observations.
We cannot see any systematic velocity gradient in the peak velocity map along the main SiO ridge.

\subsection{Position velocity diagrams}\label{PV diagrams}

\begin{figure}[htbp]
\includegraphics*[bb= 50 220 600 750, scale=0.5]{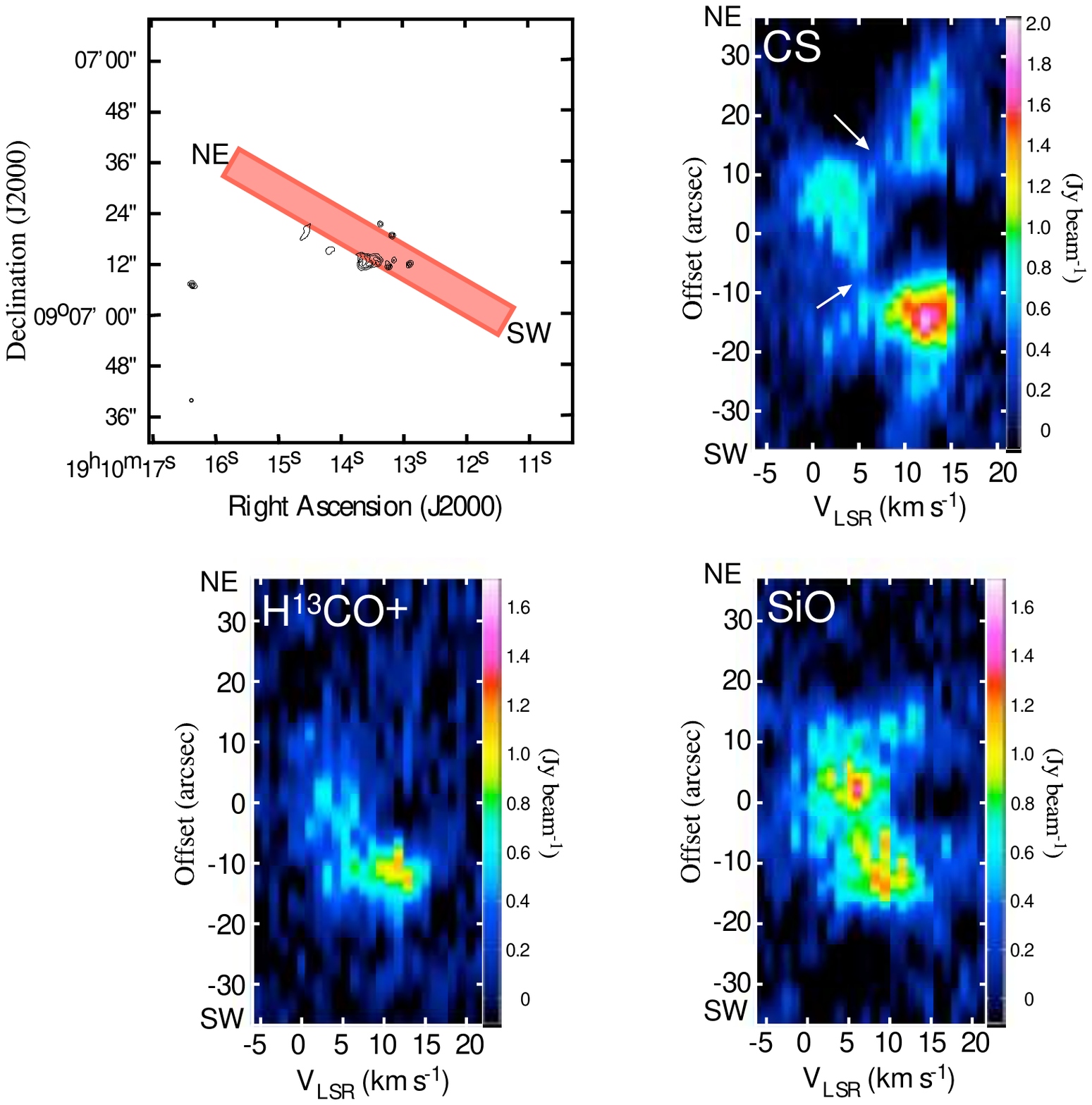}
\caption{Position-velocity (PV) diagrams along the major axis of the UCHII ring. 
{\it Upper left:} PV strip with the position angle of 60$^\circ$. The averaging slit width is 10$''$ for CS and 8$''$ for H$^{13}$CO$^+$ and SiO.
The positive direction of offsets corresponds to increasing right ascension.
{\it Upper right:} CS emission.
{\it Lower left:} H$^{13}$CO$^+$ emission.
{\it Lower right:} the SiO emission.
The arrows in the CS panel indicate the ``bridge'' features mentioned in the text.
\label{Fig07}}
\end{figure}

\begin{figure}[htbp]
\includegraphics*[bb= 50 220 600 750, scale=0.5]{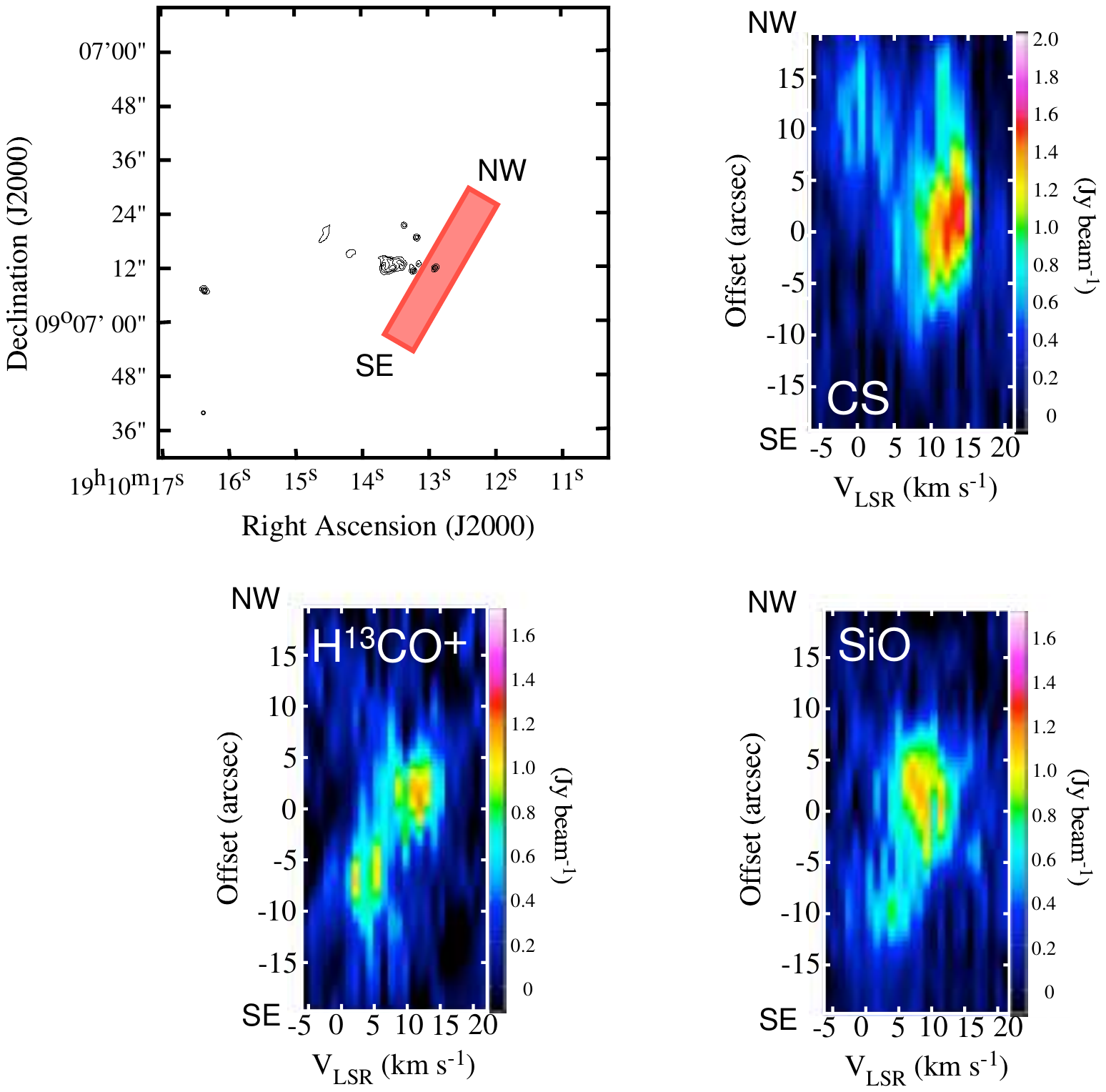}
\caption{Position-velocity (PV) diagrams through source~A along a strip parallel to the minor axis of the UCHII ring.
The positive direction of offsets correspond to increasing declination.
{\it Upper left:} PV strip with the position angle of $-$30$^\circ$. The averaging slit width is 10$''$ for CS and 8$''$ for H$^{13}$CO$^+$ and SiO.
{\it Upper right:} CS emission.
{\it Lower left:} H$^{13}$CO$^+$ emission.
{\it Lower right:} SiO emission.
\label{Fig08}}
\end{figure}

We show position-velocity (PV) diagrams in Figures~\ref{Fig07} and \ref{Fig08} in order to see that the complicated velocity field of clumps basically consists of two predominant velocity components at $\sim$4~km\,s$^{-1}$ and $\sim$12~km\,s$^{-1}$.
In Appendix~\ref{line profiles} we defined the 4~km\,s$^{-1}$ and 12~km\,s$^{-1}$ components from the line profiles of various molecular lines.

Figure~\ref{Fig07} shows the PV diagrams of the CS, H$^{13}$CO$^+$ and SiO emissions along the major axis (position angle = 60$^\circ$) of the UCHII ring.
A single feature of the 4~km\,s$^{-1}$ component and two features of the 12~km\,s$^{-1}$ component are seen well separated in CS, while the northeast feature (offset$\sim20''$) of the 12~km\,s$^{-1}$ component is not seen in H$^{13}$CO$^+$. 
The SiO emission shows two main features: one (offset$\sim0''$) positionally coincides with  the 4~km\,s$^{-1}$ component and the other (offset$\sim-10''$) with the southwestern 12~km\,s$^{-1}$ component.
The SiO velocities are slightly different from the corresponding CS or H$^{13}$CO$^+$ velocities: they are between the two characteristic velocities (4~km\,s$^{-1}$ and 12~km\,s$^{-1}$) of CS and H$^{13}$CO$^+$ emission.

Figure~\ref{Fig08} shows the PV diagrams along a strip parallel to the minor axis of the UCHII ring.
The PV strip is taken to be through source~A at a position angle of $-$30$^\circ$.
The three spectral lines show similar emission features around source~A (offset$\sim0''$) corresponding to the 12~km\,s$^{-1}$ component, although the SiO velocity is more blueshifted than those of the other two molecular species.
For the 4~km\,s$^{-1}$ component, CS has significant emission only at the northwest (offset$\sim10''$) of source~A, while both H$^{13}$CO$^+$ and SiO have emission mainly at the southeast (offset$\sim-10''$), corresponding to the clumps H$^{13}$CO$^+$-S4 and SiO-S, respectively.

\subsection{Clumps and their properties}\label{Clump properties}

\subsubsection{Identifying clumps}

We named prominent features in the velocity channel maps of CS, H$^{13}$CO$^+$ and SiO emissions in \S\ref{Result-CS}-\S\ref{Result-SiO} to describe the distribution and kinematics of the ``clumps.''
We should note that the identification of clumps is partly arbitrary.
This kind of arbitrariness particularly arises when we try to identify weak, diffuse features with apparently multiple peaks such as CS-NE12.
We tried to employ a quantitative method to identify clumps, but did not obtain satisfactory results.
The mass of all clumps added, for example, became systematically larger than the mass calculated from the spatially integrated line emission on a map.
An extensive study of clump identification algorithms shows that most of them exhibit significant deviation in extracting the total flux of clumps \citep{Li2020}.
We thus decided to use eye inspection to identify clumps with the caveat that our clump identification is partly arbitrary.
We will not go into details of the size and mass spectra of the clumps, but will simply discuss their mass and density characteristics, so that we can place reliance on the derived results.

In the end, we identified 11 CS, 8 H$^{13}$CO$^+$, and 6 SiO clumps.
Their properties are summarized in Table~\ref{Table02}, where $I_{\rm peak}$ is the peak specific intensity with a typical calibration error of 10\%, $\Delta V$ is the full velocity width at half intensity, and $\int T_{\rm MB}\,dV$ is the integrated intensity calculated from $I_{\rm peak}$ and $\Delta V$.
The radius $r$ is the geometrical average of the semimajor and semiminor axes.
Other parameters will be explained later.

The velocity width $\Delta V$(CS) ranges from 2.8~km s$^{-1}$ to 7.9~km\,s$^{-1}$ for the CS clumps, with the mean velocity width $\overline{\Delta V}$(CS) being 5.17$\pm$0.46~km\,s$^{-1}$.
For the H$^{13}$CO$^+$ clumps, the velocity width range and mean velocity width are $\Delta V({\rm H}^{13}{\rm CO}^+)=4.7-14.2$~km\,s$^{-1}$ and $\overline{\Delta V}({\rm H}^{13}{\rm CO}^+)=8.79\pm 1.61$ km s$^{-1}$, respectively. 
For the SiO clumps, they are $\Delta V({\rm SiO})=7.7-14.0$ km s$^{-1}$ and $\overline{\Delta V}({\rm SiO})=10.68\pm1.03$ km s$^{-1}$, respectively.  

\small
\begin{table*}[tp]
\begin{minipage}{\textwidth}
\caption{Clumps and their properties}
\label{Table02} 
\begin{center}
\scalebox{0.62}[0.62]{
\begin{tabular}{lccccccccccccccc}
\hline
\hline
Clump & $\alpha$(J2000) & $\delta$(J2000) & $I_{\rm peak}$ & $V_{\rm LSR}$ & $\Delta V$ & $\int T_{\rm MB}\,dV$ & Radius & $N$(mol) & $N$(H$_2$) & $n$(H$_2$) & $M$ & $M_{\rm vir}$ & $M/M_{\rm vir}$ & $t_{\rm ff}$ \\

Name  & 19$^{\rm h}$10$^{\rm m}$+($^{\rm s}$)& 9$^\circ$6$'$+($''$)& (mJy\,beam$^{-1}$) & (km\,s$^{-1}$) & (km\,s$^{-1}$) & (K\,km\,s$^{-1}$) & (pc) & (cm$^{-2}$) & (cm$^{-2}$) & (cm$^{-3}$) & (M$_\odot$) & (M$_\odot$) &  & (yr) \\
\hline

CS-C4    & 13.56 & 19.3    & 834  & 3.0   & 7.89 & 56.0 & 0.52 & 4.53E+15 & 2.06E+24 & 6.42E+05 & 2.75E+04 & 2.04E+04 & 4.05 & 6.29E+04 \\
CS-N4    & 13.56 & 34.3    & 849  & 5.9   & 7.84 & 56.6 & 0.45 & 4.58E+15 & 2.08E+24 & 7.50E+05 & 2.09E+04 & 1.74E+04 & 3.59 & 5.82E+04 \\
CS-NW4   & 12.54 & 15.3    & 824  & 3.6   & 6.05 & 42.4 & 0.27 & 3.43E+15 & 1.56E+24 & 9.26E+05 & 5.74E+03 & 6.29E+03 & 2.74 & 5.24E+04 \\
CS-SW4   & 12.85 & 9.3     & 905  & 5.5   & 4.34 & 33.6 & 0.22 & 2.72E+15 & 1.23E+24 & 9.28E+05 & 2.84E+03 & 2.56E+03 & 3.33 & 5.23E+04 \\
CS-NE12  & 14.71 & 28.3    & 926  & 12.0  & 4.84 & 38.2 & 0.37 & 3.08E+15 & 1.40E+24 & 6.06E+05 & 9.77E+03 & 5.53E+03 & 5.30 & 6.47E+04 \\
CS-SE12A & 15.63 & 1.4     & 718  & 12.0  & 3.45 & 21.0 & 0.16 & 1.70E+15 & 7.73E+23 & 7.61E+05 & 1.04E+03 & 1.23E+03 & 2.52 & 5.78E+04 \\
CS-SE12B & 14.41 & $-$1.1  & 718  & 13.4  & 4.17 & 25.5 & 0.14 & 2.06E+15 & 9.38E+23 & 1.10E+06 & 8.92E+02 & 1.52E+03 & 1.76 & 4.81E+04 \\
CS-NW12  & 11.90 & 14.3    & 1050 & 12.2  & 6.10 & 54.5 & 0.24 & 4.40E+15 & 2.00E+24 & 1.38E+06 & 5.49E+03 & 5.51E+03 & 2.99 & 4.29E+04 \\
CS-SW12  & 12.81 & 11.8    & 1740 & 11.9  & 6.23 & 93.9 & 0.37 & 7.59E+15 & 3.45E+24 & 1.50E+06 & 2.36E+04 & 9.09E+03 & 7.80 & 4.11E+04 \\
CS-S12   & 11.90 & $-$19.1 & 1006 & 11.3  & 2.80 & 24.0 & 0.20 & 1.94E+15 & 8.82E+23 & 6.99E+05 & 1.83E+03 & 1.01E+03 & 5.43 & 6.03E+04 \\
CS-SE0   & 16.68 & $-$11.1 & 892 & $-$1.3 & 3.11 & 27.7 & 0.26 & 2.24E+15 & 1.02E+24 & 6.27E+05 & 3.48E+03 & 5.34E+02 & 6.53 & 6.37E+04 \\
&  &  &  &  &  &  &  &  &  &  &  &  &  &  \\
H$^{13}$CO$^+$-S4   & 14.94 & $-$12.0 & 427  & 3.1  & 8.40  & 35.1 & 0.15 & 4.72E+13 & 1.63E+24 & 1.76E+06 & 1.81E+03 & 6.65E+03 & 0.82 & 3.72E+04 \\
H$^{13}$CO$^+$-C4   & 13.86 & 27.4    & 210  & 2.0  & 6.58  & 13.5 & 0.12 & 1.82E+13 & 6.28E+23 & 8.54E+05 & 4.42E+02 & 3.25E+03 & 0.41 & 5.34E+04 \\
H$^{13}$CO$^+$-N4   & 13.56 & 9.4     & 1288 & 3.1  & 6.36  & 80.2 & 0.35 & 1.08E+14 & 3.73E+24 & 1.72E+06 & 2.28E+04 & 8.95E+03 & 7.64 & 3.76E+04 \\
H$^{13}$CO$^+$-SE4  & 14.20 & 10.4    & 292  & 5.8  & 9.62  & 27.6 & 0.21 & 3.72E+13 & 1.28E+24 & 9.73E+05 & 2.90E+03 & 1.25E+04 & 0.70 & 5.00E+04 \\
H$^{13}$CO$^+$-SW12 & 12.98 & 12.4    & 714  & 10.0 & 14.20 & 98.9 & 0.42 & 1.33E+14 & 4.59E+24 & 1.79E+06 & 3.94E+04 & 5.28E+04 & 2.24 & 3.69E+04 \\
H$^{13}$CO$^+$-NW12 & 12.10 & 15.9    & 329  & 11.2 & 4.72  & 15.2 & 0.16 & 2.05E+13 & 7.05E+23 & 7.11E+05 & 9.03E+02 & 2.26E+03 & 1.20 & 5.85E+04 \\
H$^{13}$CO$^+$-SE12 & 14.20 & 4.5     & 212  & 10.5 & 12.20 & 25.3 & 0.11 & 3.41E+13 & 1.18E+24 & 1.72E+06 & 7.16E+02 & 1.04E+04 & 0.21 & 3.77E+04 \\
H$^{13}$CO$^+$-S12  & 13.42 & $-$3.5  & 299  & 7.0  & 8.26  & 24.2 & 0.14 & 3.26E+13 & 1.12E+24 & 1.31E+06 & 1.07E+03 & 5.96E+03 & 0.54 & 4.31E+04 \\
 &  &  &  &  &  &  &  &  &  &  &  &  &  &  \\

SiO-C  & 13.62 & 18.4 & 1233 & 4.7  & 9.20  & 111.3 & 0.30 & 4.06E+14 & 1.07E+25 & 5.70E+06 & 4.88E+04 & 1.62E+04 & 9.05 & 1.25E+04 \\
SiO-NE & 14.17 & 24.4 & 710  & 10.9 & 13.40 & 93.2  & 0.15 & 3.40E+14 & 8.95E+24 & 9.89E+06 & 9.54E+03 & 1.66E+04 & 1.73 & 9.52E+03 \\
SiO-SW & 12.64 & 11.4 & 1118 & 9.0  & 10.26 & 112.3 & 0.26 & 4.10E+14 & 1.08E+25 & 6.61E+06 & 3.74E+04 & 1.75E+04 & 6.39 & 1.16E+04 \\
SiO-SE & 15.48 & 10.3 & 580  & 8.7  & 7.70  & 43.7  & 0.15 & 1.59E+14 & 4.20E+24 & 4.54E+06 & 4.66E+03 & 5.59E+03 & 2.50 & 1.40E+04 \\
SiO-S  & 19.45 & 5.4  & 846  & 3.4  & 9.50  & 79.1  & 0.14 & 2.89E+14 & 7.60E+24 & 8.88E+06 & 7.23E+03 & 7.88E+03 & 2.75 & 1.00E+04 \\
SiO-W  & 21.53 & 15.8 & 915  & 7.9  & 13.99 & 125.1 & 0.19 & 4.57E+14 & 1.20E+25 & 1.00E+07 & 2.24E+04 & 2.39E+04 & 2.81 & 9.45E+03 \\

\hline
%
\vspace{3mm}

\end{tabular}
}
\end{center}
\footnotetext{
}
\end{minipage}
\end{table*}

\normalsize


The radii of the clumps are $r$(CS) = 0.14 -- 0.52~pc, $r$(H$^{13}$CO$^+$) = 0.11 -- 0.42~pc, and $r$(SiO) = 0.14 -- 0.30~pc, with their mean values of $\overline{r}$(CS) = 0.29$\pm$0.12~pc, $\overline{r}$(H$^{13}$CO$^+$ ) = 0.21$\pm$0.11~pc, and $\overline{r}$(SiO) = 0.07$\pm$0.11~pc, respectively. 
The minimum radii are close to the beam radii of $\sim$0.22~pc and $\sim$~0.11~pc for 49~GHz and 86~GHz observations, respectively, implying that some of the clumps are not resolved along the minor axes.
The two CS features CS-SE12A and CS-SE12B have significantly small radii compared with that of the 49~GHz beam.
These are weak emission features as seen in Figure~\ref{Fig01} and may have been detected only at and around their peaks, hampering us from accurately measuring the half maximum size, which probably resulted in the apparently small radii.

\begin{figure}[htbp]
\includegraphics*[bb= 120 100 600 750, scale=0.7]{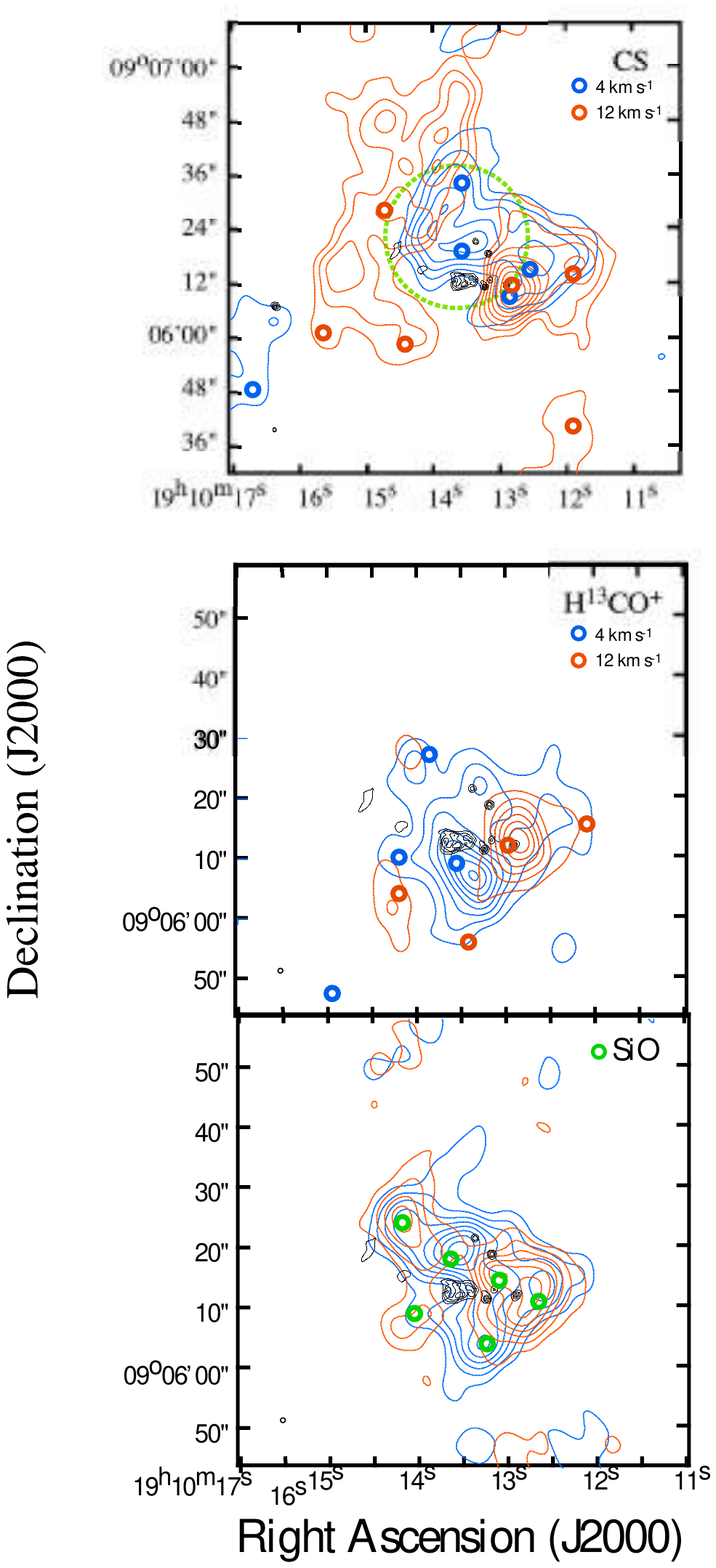}
\caption{Peak positions of the clumps are plotted on the contour maps of the CS ({\it Top}), H$^{13}$CO$^+$ ({\it Center}), and SiO ({\it Bottom}) emissions.
For the CS and H$^{13}$CO$^+$ plots, the position markers ({\it open circles}) are color-coded as blue and red for the 4~km\,s$^{-1}$ and 12~km\,s$^{-1}$ components, respectively.
Blue and red contours indicate the intensities integrated over $V_{\rm LSR}\leq\,$8~km\,s$^{-1}$ (4~km\,s$^{-1}$ component) and $V_{\rm LSR}\geq\,$8~km\,s$^{-1}$ (12~km\,s$^{-1}$ component), respectively.
The 8.3~GHz continuum map \citep{DePree1997} is also shown in black contours.
{\it Green dotted line circle} in the {\it top panel}, roughly coinciding with the 4\,km\,s$^{-1}$ component of CS, indicates what we think the hole of the 12\,km\,s$^{-1}$ component (see text).
\label{Fig09}}
\end{figure}

Figure~\ref{Fig09} shows the locations of the clumps superposed on the contour maps of CS, H$^{13}$CO$^+$, and SiO emissions integrated over $\leq\,$8~km\,s$^{-1}$ (blue) and $\geq\,$8~km\,s$^{-1}$ (red).
For the CS and H$^{13}$CO$^+$ plots, the plot markers are also color-coded as blue and red for the 4~km\,s$^{-1}$ and 12~km\,s$^{-1}$ components, respectively.

The top panel of Figure~\ref{Fig09} shows the distribution of CS clumps, which is naturally consistent with the contour maps: the 4~km\,s$^{-1}$ clumps are concentrated around the UCHII ring, while the 12~km\,s$^{-1}$ clumps are on the elongated feature at the east of the UCHII ring and at the west around source~A.
The blueshifted clump CS-SE0 is at the lower left corner.
The center panel shows the distribution of H$^{13}$CO$^+$ clumps. 
The three 4~km\,s$^{-1}$ clumps are located toward the UCHII ring, and the 12~km\,s$^{-1}$ clumps are at the west and south of them.
The bottom panel shows the distribution of SiO clumps. 
The clumps are concentrated in and around the UCHII ring.

\subsubsection{Column density of the clumps}\label{Column density}

We calculated the column density $N$(mol) of CS, H$^{13}$CO$^+$ and SiO toward the peaks of the clumps as listed in Table~\ref{Table02}.
It is calculated from the integrated intensity ($\int T_{\rm MB}\,dV \,\, \mathrm{[K\,km\, s^{-1}]}$) of the line emission under the assumption that all the rotational levels are populated with the same excitation temperature $T_{\rm ex}$ specific to the molecular species:
\begin{eqnarray}
\mathrm{{\it N}(mol)\,\,[cm^{-2}]} &=& \mathrm{\frac{1.94 \times 10^3}{2\it{J}+3} \frac{{\nu_{0}^3}}{{\it B}\,{\it A}_{\it J+1, J}} \frac{\mathrm{exp}\left[{-0.048\,\it{B}\,J(J+1)}/{\it{T}_{\rm ex}}\right]}{1-\mathrm{exp}\left[{-0.048\,\nu_0}/{\it{T}_{\rm ex}}\right]}} \nonumber \\
&\times&\frac{\tau}{1-\mathrm{exp}({-\tau})} \int \it{T}_{\rm MB}\,dV,
\end{eqnarray}

\noindent
where $\nu_0$ [GHz], $B$ [GHz] and $A_{J+1, J}$ [s$^{-1}$] are the frequency, rotational constant, and Einstein coefficient, respectively, of the observed transitions of CS ($J=1-0$), H$^{13}$CO$^+$ ($J=1-0$) and SiO ($J =2-1$) (see  Table~\ref{Table01}).

The factor ${\tau}/[1-\mathrm{exp}({-\tau})]$ is the opacity correction for optically thick emission.
For H$^{13}$CO$^+$ ($J=1-0$), this factor is essentially 1 because the intensity ratio of $T_{\rm A}($H$^{13}$CO$^+$) / $T_{\rm A}$(H$^{12}$CO$^+$) = 0.16 (see figure~\ref{Fig17}) means $\tau$(H$^{13}$CO$^+$) = 0.15.
The optical depth of the CS ($J=1-0$) emission is assumed to be  $\tau$(CS)$\sim$5 taken from \citet{Miyawaki1986}, corresponding to the opacity correction factor of 5.
We assumed the optical depth of SiO to be $\sim$1.5, taken from $\tau$(SiO) =1.0--2.2 obtained toward a shocked region in L1157-B1 \citep{Spezzano2020}, so  the opacity correction factor is $\sim$1.9.

The excitation temperature $T_{\rm ex}$ was assumed to be 20~K for all the three observed molecular species.
The value seems to be a little high for dark clouds, but is reasonable when we consider the surface brightness of the line emissions.
The maximum specific intensities for CS, H$^{13}$CO$^+$ and SiO are 2.0, 1.5 and 1.6 Jy\,beam$^{-1}$, respectively, on their respective maps.
With the specific intensity to brightness temperature conversion factor of 8.02~K\,(Jy\,beam$^{-1}$)$^{-1}$ at 49.0 GHz for the 9\farcs2$\times$ 6\farcs9 beam and 9.33~K\,(Jy\,beam$^{-1}$)$^{-1}$ at 86.7~GHz for the 4\farcs7$\times\,$3\farcs7 beam, we obtain the excess brightness temperature of 14--16~K for all the three lines.
When we add the contribution from the cosmic microwave background radiation, the corresponding brightness temperature is $\sim$18~K, which should be the lower limit to the excitation temperature.
The column density is approximately proportional to the excitation temperature for $T_{\rm ex}\,\ga\,5\,$K, so a factor of two uncertainty in the excitation temperature propagates at the same factor into the density and mass we derive below.

The column density of each molecular species was divided by the corresponding abundance ratio to obtain the H$_2$ column density $N$(H$_2$), which is listed in Table~\ref{Table02}.
We assumed the abundance ratios of 2.2$\times$$10^{-9}$, 2.9$\times$$10^{-11}$, and 3.8$\times$$10^{-11}$ for CS, H$^{13}$CO$^+$ and SiO, respectively, which were measured toward Orion and other HII regions by \citet{Li2019} and \citet{Rodriguez-Baras2021}.
The abundance ratios may be the most uncertain factor, varying an order of magnitude among the Galactic sources according to the studies these authors presented.

Examining published values of fractional abundances, we estimate a typical error in the abundances of CS and H$^{13}$CO$^+$ to be about a factor of three.
Hence the resultant errors in the molecular hydrogen column densities, masses, and densities of CS and H$^{13}$CO$^+$ clumps would be a factor of several times, when we also take into account the uncertainty in the excitation temperature.
The fractional abundance of SiO could be more uncertain.
There are studies that it may reach much larger than 10$^{-10}$ \citep{Gibb2004, Jimenez-Serra2010}.
If the SiO abundance of the presently studied areas takes a value much larger than 10$^{-10}$, the resultant molecular hydrogen column density, mass, and density of SiO clumps would be an order of magnitude smaller than the values derived in this paper.

\subsection{Mass of the clumps}\label{Mass}

The gas mass $M$ of each clump was derived from $N$(H$_2$) multiplied by the surface area of the clump with a radius of $r$. 
The range and mean of the masses derived for the 11 CS clumps are $M{\rm (CS)} = 8.9\times10^2-2.8\times10^4$  M$_{\odot}$ 
and $\overline{M}{\rm (CS)}=9.4\times10^3$ M$_{\odot}$, respectively. 
For the 8 H$^{13}$CO$^+$ clumps, the range and mean are $M{\rm (H^{13}CO^+)}=4.4\times10^2-3.9\times10^4$  M$_{\odot}$ and $\overline{M}{\rm (H^{13}CO^+)}=8.8\times10^3$ M$_{\odot}$, respectively. 
For the 6 SiO clumps, they are $M{\rm (SiO)}= 4.7\times10^3-4.9\times10^4$  M$_{\odot}$ and $\overline{M}{\rm (SiO)}=2.2\times10^4$ M$_{\odot}$, respectively. 
The total masses for the CS, H$^{13}$CO$^+$, and SiO clumps are 1.0$\times$10$^5$~M$_{\odot}$, 7.0$\times$10$^4$~M$_{\odot}$, and 1.3$\times$10$^5$~M$_{\odot}$, respectively.
These values are consistent with the total mass ($\sim10^5$~M$_{\odot}$) of the W49N core measured with single dish telescopes.
We should not add the masses obtained from the three emission lines, because some of the clumps detected in different lines are physically the same and some are not.

We also estimated the mass $M_{\rm vir}$ of each clump using the virial theorem assuming no external pressure and no magnetic field. 
This ``virial'' mass was calculated from the radius $r$ and line width $\Delta V$ of the clumps as $M_{\rm vir} = 3\,r\,\Delta V^2/G$, where the factor 3 converts the one dimensional (line of sight) velocity dispersion into the three dimensional value.
The range and mean of the virial masses for the CS clumps are $M_{\mathrm{vir}}{\rm (CS)}= 5.3\times10^2-2.0\times10^4$  M$_{\odot}$ and $\overline{M}_{\mathrm{vir}}{\rm (CS)}=6.5\times10^3$~M$_{\odot}$, respectively. 
The range and mean for the H$^{13}$CO$^+$ clumps are  $M_{\mathrm{vir}}{\rm (H^{13}CO^+)}= 2.3\times10^3-5.3\times10^4$  M$_{\odot}$ and $\overline{M}_{\mathrm{vir}}{\rm (H^{13}CO^+)}=1.3\times10^4$~M$_{\odot}$, respectively. 
For the SiO clumps, they are  $M_{\mathrm{vir}}{\rm (SiO)}= 5.6\times10^3-2.4\times10^4$  M$_{\odot}$ and $\overline{M}_{\mathrm{vir}}{\rm (SiO)}=1.5\times10^4$~M$_{\odot}$, respectively. 
The total virial masses for the CS, H$^{13}$CO$^+$, and SiO clumps are 7.1$\times$10$^4$~M$_{\odot}$, 1.3$\times$10$^5$~M$_{\odot}$, and 8.8$\times$10$^4$~M$_{\odot}$, respectively.

\begin{figure}
\includegraphics*[bb= 0 150 550 600, scale=0.40]{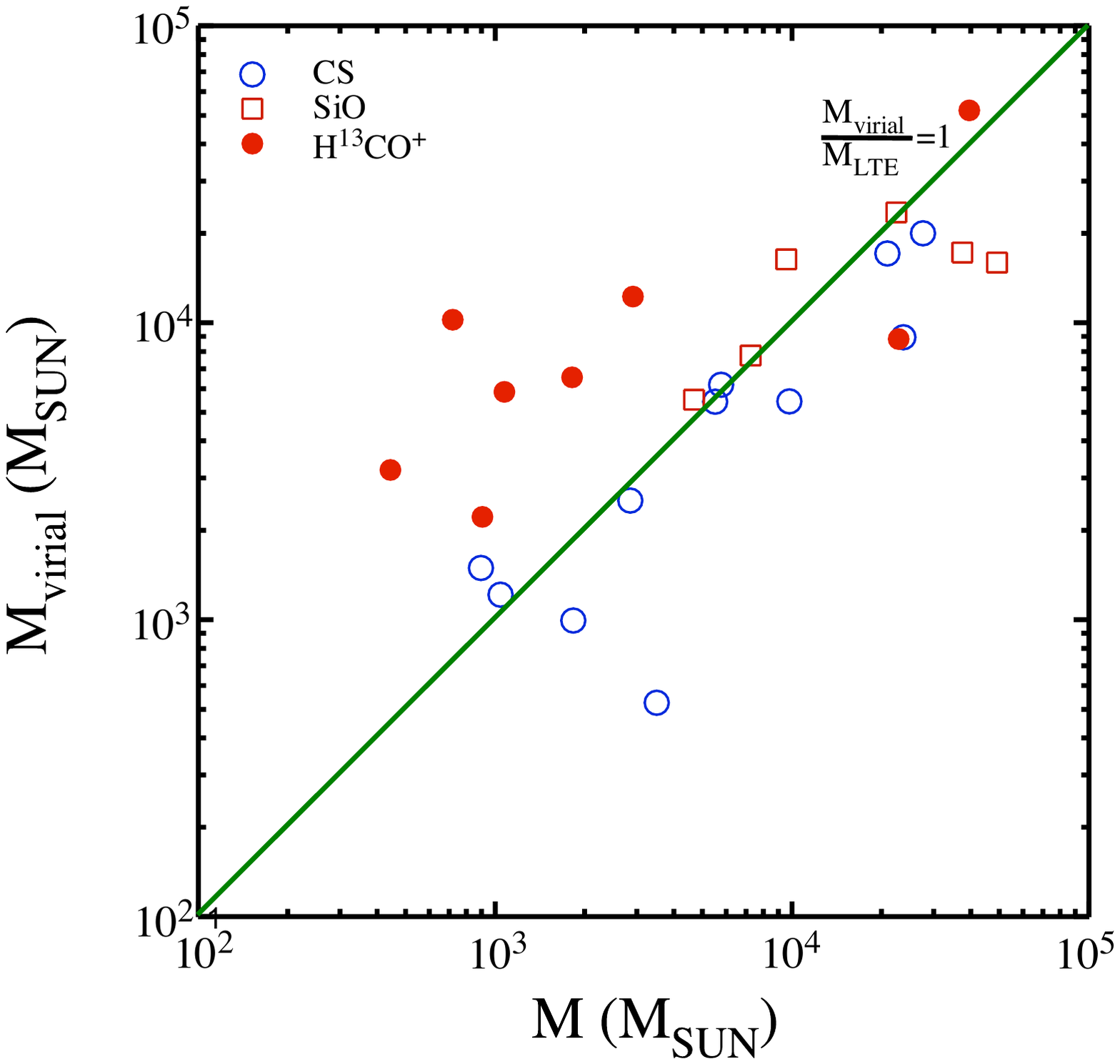}
\caption{
Clump mass - virial mass relation for the CS ({\it blue open circles}), SiO ({\it red open squares}), and H$^{13}$CO$^+$ ({\it red filled circles}) clumps.
\label{Fig10}}
\end{figure}

Figure~\ref{Fig10} shows the relation between the gas mass $M$ and virial mass $M_{\rm vir}$ of the clumps.
The mean values of gas mass to virial mass ratios, listed in Table~\ref{Table02}, are 1.8, 0.57 and 1.4 for the CS, H$^{13}$CO$^+$ and SiO clumps, respectively.
The two types of mass agree well with each other, when we consider the typical error of a factor of several times caused by the uncertainties in the molecular abundances and excitation temperature.

\begin{figure}
\includegraphics*[bb= 0 150 500 620, scale=0.4]{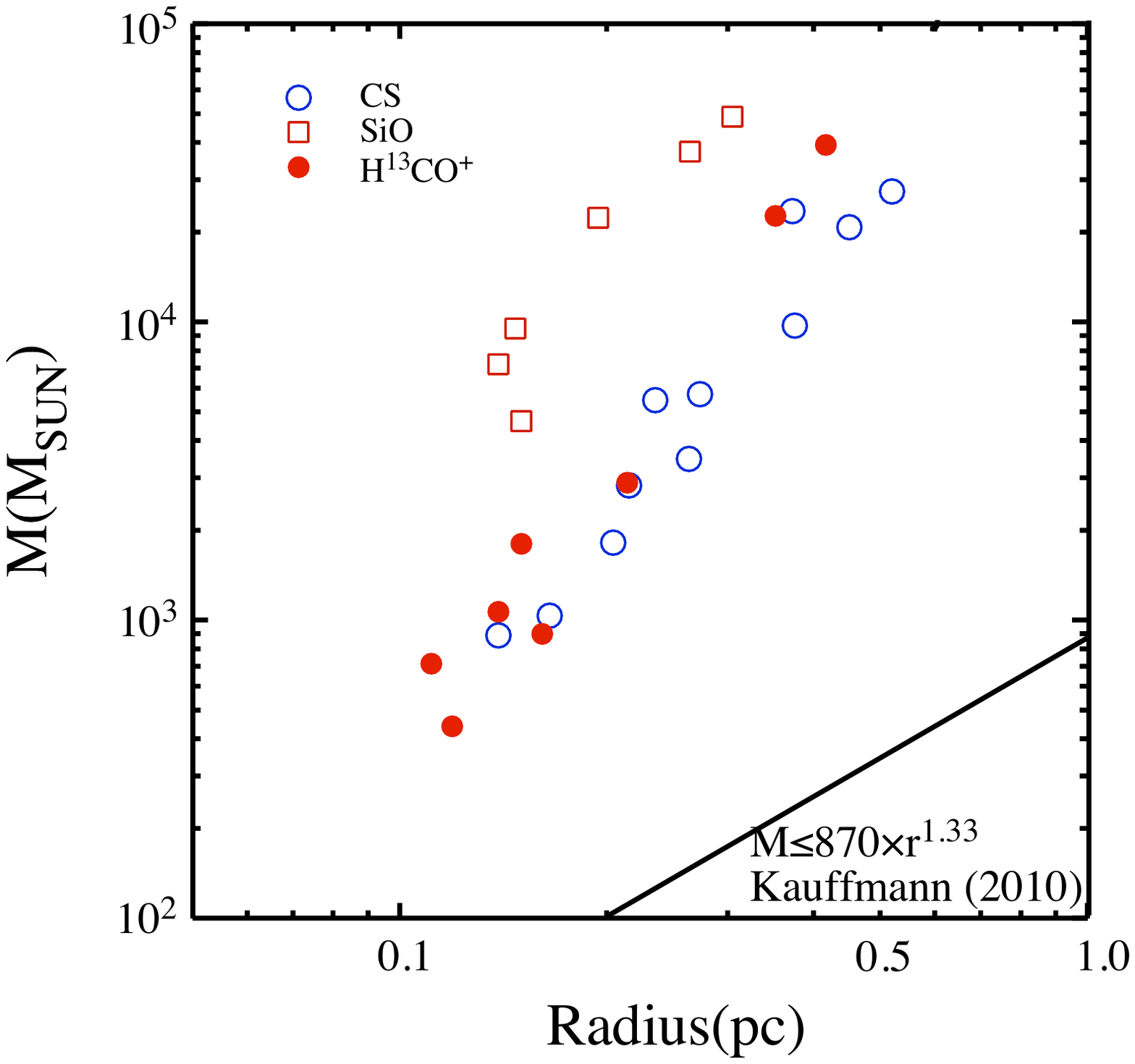}
\caption{
Gas mass - radius relation for the CS ({\it blue open circles}), SiO ({\it red open squares}), and H$^{13}$CO$^+$ ({\it red filled circles}) clumps.
Solid line shows the threshold for massive star formation proposed by \citet{Kauffmann2010}. 
\label{Fig11}}
\end{figure}

Figure~\ref{Fig11} shows the relation between the gas mass $M$ and radius of the clumps.
There is a clear trend that the mass is proportional to $r^3$, although a systematic difference is visible between the SiO and other clumps.
All the clumps are located above the line representing $M= 870\,r^{1.33},$ which divides cloud fragments with and without massive star formation.
The relation, similar to the one originally discovered by \citet{Larson1981}, was derived by \citet{Kauffmann2010} from the mass and size of cloud fragments over a wide range of spatial scales (0.05 $\la\,r\,\la$ 3 pc).
The result that all the W49N clumps are located above this line means that they have characteristics of massive star formation, different from those in low mass star forming regions.

\subsubsection{Density of the clumps}\label{Density}

\begin{figure}
\includegraphics*[bb= 0 150 500 620, scale=0.4]{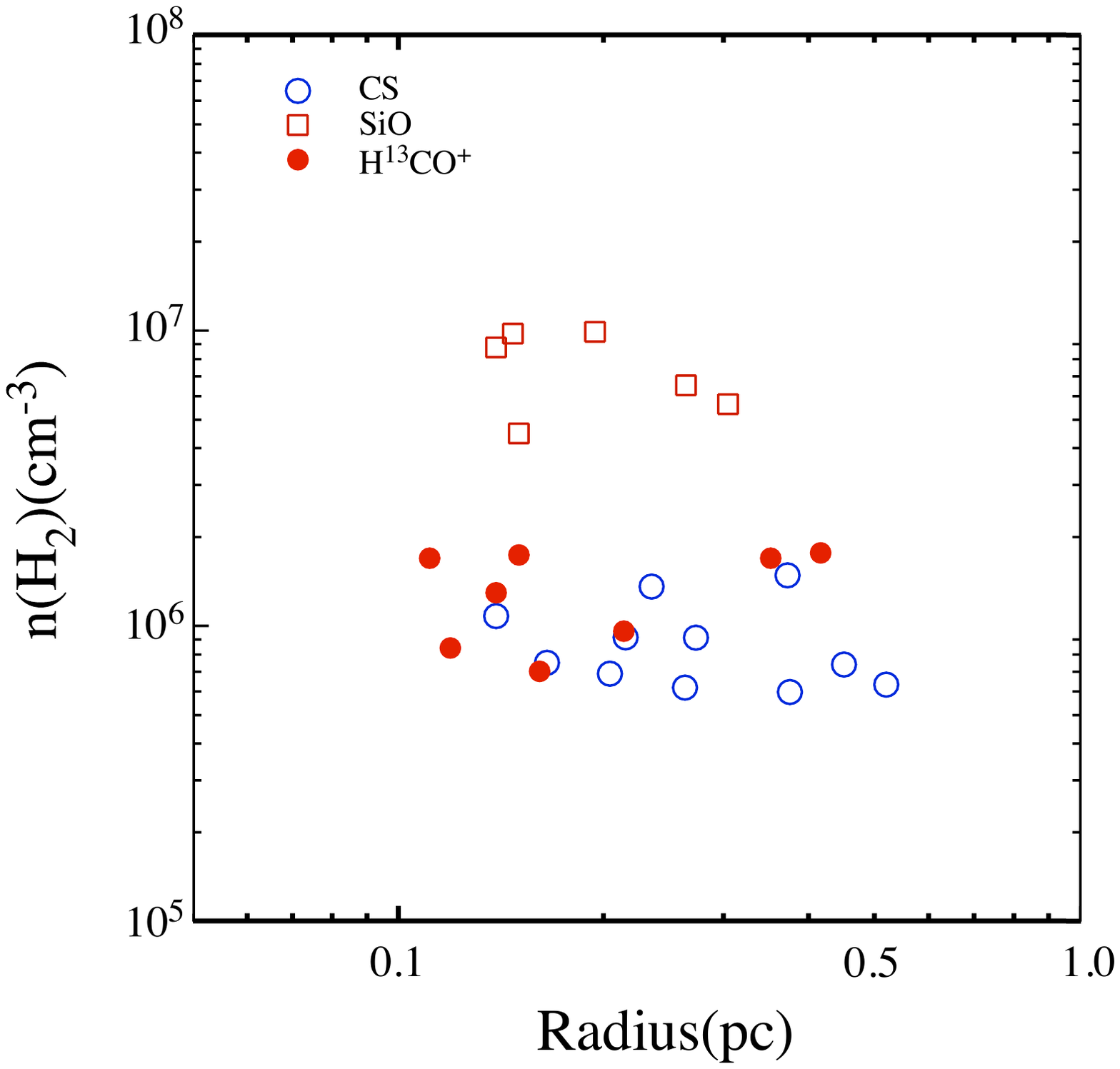}
\caption{
Number density - size relation of the CS ({\it blue open circles}), SiO ({\it red open squares}), and H$^{13}$CO$^+$ ({\it red filled circles}) clumps.
\label{Fig12}}
\end{figure}

Figure~\ref{Fig12} shows the relation between the molecular hydrogen number density $n({\rm H_2}),$ calculated from $N({\rm H_2})$ divided by $2r$.
We do not see a clear tendency that the density varies with radius.
The average densities are $9.0\times 10^5\,$cm$^{-3}$, $1.4\times 10^6\,$cm$^{-3}$, and $7.6\times 10^6\,$cm$^{-3}$ for the CS, H$^{13}$CO$^+$ and SiO clumps, respectively.
Because the critical densities of the observed transitions are not much different from each other, the significantly high density derived from the SiO emission may reflect the intrinsically high density of the SiO emitting regions, as is consistent with the emission arising from shocked hot cores.
We should also note that these values have errors of a factor of several times.

\subsubsection{Free fall time and mass accretion rate}
\label{Time scale}

\begin{figure}
\includegraphics*[bb= 0 150 500 620, scale=0.4]{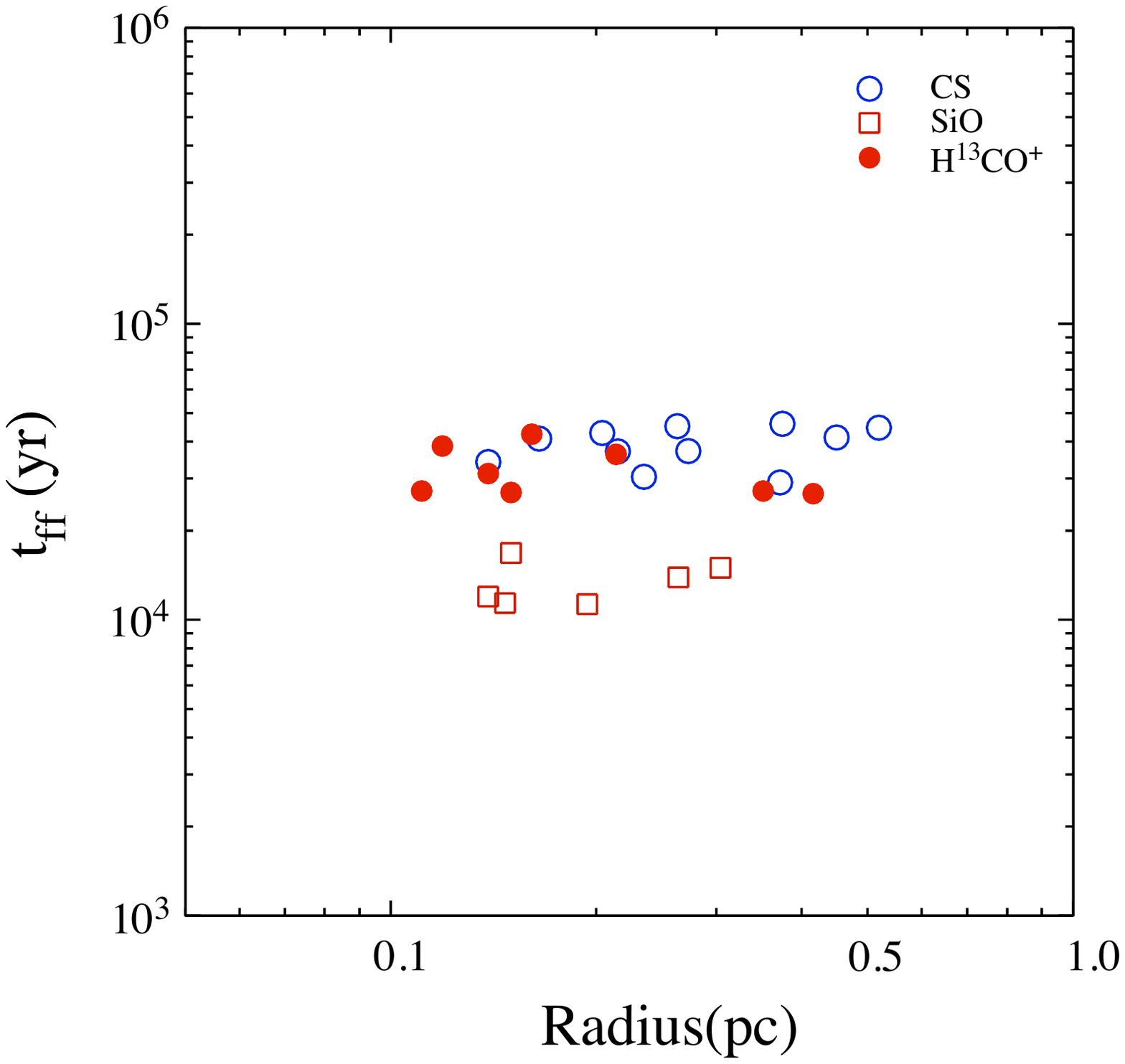}
\caption{
Free fall time - size relation of the CS ({\it blue open circles}), SiO ({\it red open squares}), and H$^{13}$CO$^+$ ({\it red filled circles}) clumps.
\label{Fig13}}
\end{figure}

The free fall time scale $t_{\rm ff}$ of each clump is calculated from the number density and is plotted against the clump radius in Figure~\ref{Fig13}.
Note that the typical error in the free fall time scale is less than a factor of three because it is proportional to the inverse square root of density.
The mean free fall time is $4.0\times 10^4$~yr, $3.3 \times 10^4$~yr, and $1.4\times 10^4$~yr for the CS, H$^{13}$CO$^+$ and SiO clumps, respectively.
The time scale is shorter than the typical free fall time scale of low mass star formation, i.e., a few times 10$^5$\,yr to 10$^6$\,yr, and is suitable for massive star formation \citep[e.g.,][]{Hosokawa2009}.
The free fall time scales of the SiO clumps are significantly smaller than those of the other molecular clumps, which may reflect the SiO emission produced in shocked regions, although we cannot exclude the possibility that systematic errors in abundances that we used to derive the column densities are the cause of this difference.

\begin{figure}
\includegraphics*[bb= 0 150 550 650, scale=0.4]{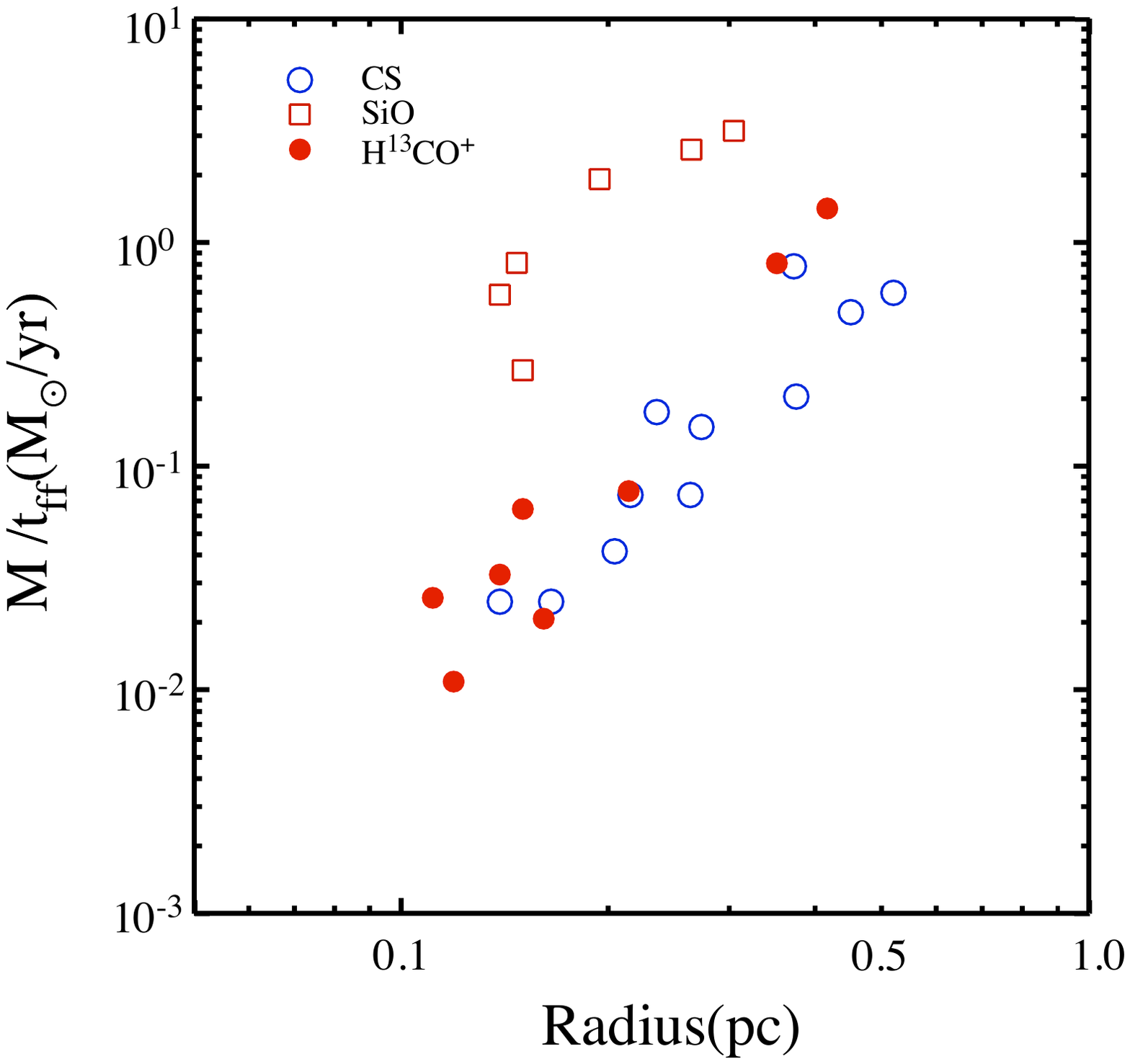}
\caption{
Nominal mass accretion rate - radius relation of the CS ({\it blue open circles}), SiO ({\it red open squares}), and H$^{13}$CO$^+$ ({\it red filled circles}) clumps.
\label{Fig14}}
\end{figure}

We may derive the ``nominal" mass accretion rate at the scale of observed clumps by dividing the gas mass $M$ by the free fall time.
The accretion rate is plotted against the clump radius in Figure~\ref{Fig14}.
It scatters over 2.5 orders of magnitude from 10$^{-2}$ to $\sim 4\,M_\odot\,{\rm yr}^{-1}$ with a tendency of rapid increase with radius.
This nominal accretion rate must be larger than the mass accretion rate onto a stellar surface from its surroundings, when we consider that a significant fraction of accreting mass is converted to an outflow near the central star.
The actual mass accretion rate onto a star $\dot{M}$ [M$_\odot\,{\rm yr}^{-1}$] is empirically calculated from $t_{\rm ff}$ [yr] as,
\begin{equation}
\dot{M} = f_{\rm infall}\,\frac{M}{t_{\rm ff}},
\end{equation}
where $f_{\rm infall}$ is the fraction of mass actually accreting onto the star out of the entire infalling mass in its envelope.
The value of $f_{\rm infall}$ is estimated to be 0.2--0.3 \citep[e.g.,][]{Behrend2001, Fazal2008, Haemmerle2016}, which means that the mass accretion rate onto a possibly embedded stellar core may be $\dot{M}=3\times10^{-3}-1\,{\rm M}_\odot\,$yr$^{-1}$ for the W49N clumps.
Previous studies show that the accretion rate $\dot{M}$ should be larger than $3\,\times\,10^{-3}~{\rm M}_\odot\,$yr$^{-1}$ for massive stars to form \citep{Wyrowski2012}.
We may thus conclude that most of the clumps identified in the present study are suitable for massive star formation.

\section{Discussion}\label{Discussion}

\subsection{Individual Sources}\label{Individual}

In this subsection, we will summarize star formation activities of individual UCHII regions and hot molecular cores referring to previous observations in connection with the present results.
Maps of W49N in many {\it hot core} molecular lines are available in \citet{Madrid2013}.
They presented high resolution ($\sim1''$) maps obtained with Submillimeter Array (SMA) in the lines of HC$_3$N, CH$_3$OH, H$_2$CO, SO$_2$, OCS, HNCO, CH$_3$CH$_2$CN, SO, CH$_3$OCHO, CH$_3$CHO, CH$_3$CN, CH$_3$CCH, H$_2$CCO, $^{13}$CS, and CH$_3$OCH$_3$ at 220~GHz -- 230~GHz.
High resolution continuum emission and recombination line maps are also available at 3.6~cm, 7~mm and 1.4~mm \citep{Wilner2001, DePree2004, DePree2020}.
These observations, including the present study, have revealed a variety of sources in W49N such as the hot core MCN-a and bipolar UCHII region source~A.
The variety may reflect different phases of massive star formation in W49N.

\subsubsection{Source~A}
Source~A has a clear edge-brightened, double-lobed (bipolar) structure, with the lobes extending $\sim$1\,\farcs9 to the northwest and 0\,\farcs8 to the southeast ($\sim$0.1 pc) \citep{DePree2020}.
Its recombination line velocity is $V_{\rm LSR}$= 13.2$\pm$0.7 km\,s$^{-1}$ and 19.3$\pm$2.1 km\,s$^{-1}$ for H52$\alpha$ and H92$\alpha$, respectively \citep{DePree2020}.

The strongest emission features of all the three molecular lines we observed are toward or in the immediate vicinity of source~A, suggesting that the UCHII region is still associated with plenty of dense molecular gas.
Because SiO is a tracer of shocks often produced by outflows \citep[e.g.,][]{Lopez2016}, we may naturally assume that the SiO emission surrounding the bipolar UCHII region arises, at least part of it, in the shocked regions caused by the rapid bipolar outflow of ionized gas.
If we take a closer look at the integrated intensity map of SiO (Figure~\ref{Fig06}), source~A is located in a shallow dip of emission surrounded by SiO-S, SiO-NW and SiO-SW, as noted in the previous section. 
The bipolar UCHII outflow may have removed a significant amount of molecular gas from its surroundings.

In the molecular line maps of \citet{Madrid2013}, only $^{13}$CS ($J=5-4$), CH$_3$OCHO, and CH$_3$CCH show considerable emission around source~A.
While their $^{13}$CS emission does not show noticeable feature, CH$_3$OCHO and CH$_3$CCH maps have conspicuous emission near source A, especially at $\sim1''$ west of it.
This may correspond to SiO-S in our SiO map.

\subsubsection{Source~B}
Source~B, located between A and D, is a less conspicuous UCHII region in the 8.3~GHz map \citep{DePree1997}, but becomes prominent at 219~GHz \citep{Madrid2013}.
It is resolved into three sources B1, B2 and B3 \citep{DePree2020}, among which B2 is the brightest at 219~GHz.
Source~B1 has the H52$\alpha$ line velocity of $V_{\rm LSR}$= 6.1$\pm$2.4 km\,s$^{-1}$, while source~B2 show the recombination line velocities of $V_{\rm LSR}$= 15.9$\pm$0.8 km\,s$^{-1}$ and 7.2$\pm$1.8 km\,s$^{-1}$ for the H52$\alpha$ and H92$\alpha$ lines, respectively \citep{DePree2020}.
This source is located at the eastern edge of CS-SW12 and H$^{13}$CO$^+$-SW12, and at the southern edge of SiO-SW.
Emission lines of CH$_3$OH, HC$_3$N, HNCO, SO$_2$, CH$_3$OCHO, OCS, CH$_3$CH$_2$CN, CH$_3$CN, and CH$_3$OCH$_3$ have compact features coincident with source~B \citep{Madrid2013}, suggesting that the source is accompanied by a hot molecular core.

\subsubsection{Sources~C and F}
Sources~C and F form part of  the northern UCHII ring and are located close to or toward CS-C4. 
The recombination line velocities of source~C are $V_{\rm LSR}$= $-$9.5$\pm$0.8 km\,s$^{-1}$ and 3.4$\pm$21.4 km\,s$^{-1}$ for the H52$\alpha$ and H92$\alpha$ lines, respectively.
In spite of its rather {\it normal} radial velocity, it has a notable proper motion velocity of 76~km\,s$^{-1}$ and is interpreted as a runaway star \citep{Rodriguez2020}.
The velocity of source~F is 5.3$\pm$0.7 km\,s$^{-1}$ for the H92$\alpha$ line.

\subsubsection{Source~G}
Source~G is the brightest UCHII region at 8.3~GHz and 15~GHz \citep{Dreher1984,DePree1997} associated with strong H$_2$O masers \citep{Walker1982}.
It was resolved into many individual hyper compact HII regions: G1, G1S, G2a--c, G2ab, G3, G3a--d, G4 and G5 \citep{DePree2020}, some showing complex morphologies.
The recombination line velocities of G1, G2, and their subcomponents are $V_{\rm LSR}$= 2.4--19.9~km\,s$^{-1}$ for the H52$\alpha$ and H92$\alpha$ lines \citep{DePree2020}.

Only source~G is associated with H$_2$O masers in W49N, with the strongest H$_2$O masers concentrated on G1 and G2.
The center of maser spot expansion is located at the southern edge of G2a \citep{DePree2020}.

Emission lines of various molecules observed by \citet{Madrid2013}, including hot core molecules, have a peak toward G1 and G2, but the emission features tend to be not as conspicuous as those of MCN-a.
Source~G as a whole appears to be weak in all the three lines we observed.
The integrated intensity of SiO, for example, shows a hole toward it (Figure~\ref{Fig06}), suggesting that a cluster of O and early B stars formed there have cleared a significant amount of molecular gas, leaving relatively small amount of gas responsible for hot core and water maser emission lines.
These results suggest that source~G as a whole is in a later phase of cluster formation than the other UCHII sources and MCN-a.

\subsubsection{MCN-a}
MCN-a is a spatially compact, well isolated emission feature detected with numerous species of hot core molecules such as CH$_3$CN, CH$_3$OH, HC$_3$N, H$_2$CO, OCS, HNCO, SO$_2$, SO, CH$_3$OCHO, $^{13}$CS, CH$_3$CH$_2$CN, and CH$_3$OCH$_3$ \citep{Wilner2001, Madrid2013}.
It coincides with the 1.4 mm source K2 \citep{Wilner2001}, a dust continuum source with the flux of 200$\pm$40 mJy.
The source position agrees within an accuracy of 0\farcs3 with source~J1, an utterly inconspicuous continuum feature with a flux of 22 mJy at 3.6 cm \citep{DePree1997}.
It is not detected at higher frequencies with flux upper limits of $\la$10 mJy  \citep{DePree1997} and $<$12 mJy (3$\sigma$) \citep{DePree2000} at 1.3 cm and 3.4 mm, respectively, until it appears as the 1.4 mm source.

The dust emission, compact features of hot core molecules, and little free-free emission may place MCN-a in a very early phase of massive star formation.
Its association with SiO-NE, a relatively well defined SiO emission feature at 1.8~km\,s$^{-1}\leq V_{\rm LSR}\leq\,$13.7~km\,s$^{-1}$, implies that the earliest activities, such as outflows, of massive star formation are taking place, when we consider that SiO is a short-lived shock tracer \citep{Lopez2016}.
We may hence naturally assume that MCN-a is in an earlier phase than source-A, which is associated  with a large bipolar outflow.
Such a hot molecular core may have similar characteristics as the Galactic Center 50~km\,s$^{-1}$  Molecular Cloud \citep{Miyawaki2021}.

\subsection{A cloud-cloud collision view of W49N}
\label{CCC view}

We have seen in \S\ref{Results} that there are a few dozens of molecular gas clumps in and around the UCHII ring of the W49N core.
We did not find any systematic motion, such as rotation around the UCHII ring, of the clumps as a whole.
The clumps, with their masses totaling 10$^5$~M$_\odot$, have an average density of 10$^6$~cm$^{-3}$ and a free fall time scale of a few times 10$^4$~yr, implying that they are capable of forming another few dozens of massive stars at a mass accretion rate of $\dot{M}=3\times10^{-3}-1$~M$_\odot$\,yr$^{-1}$.
In this subsection, we will discuss these results in connection with the cloud-cloud collision view of the W49N core.

\subsubsection{Cloud-cloud collision view}\label{CCC}

\begin{figure}[htbp]
\includegraphics*[bb= 80 30 600 750, scale=0.55]{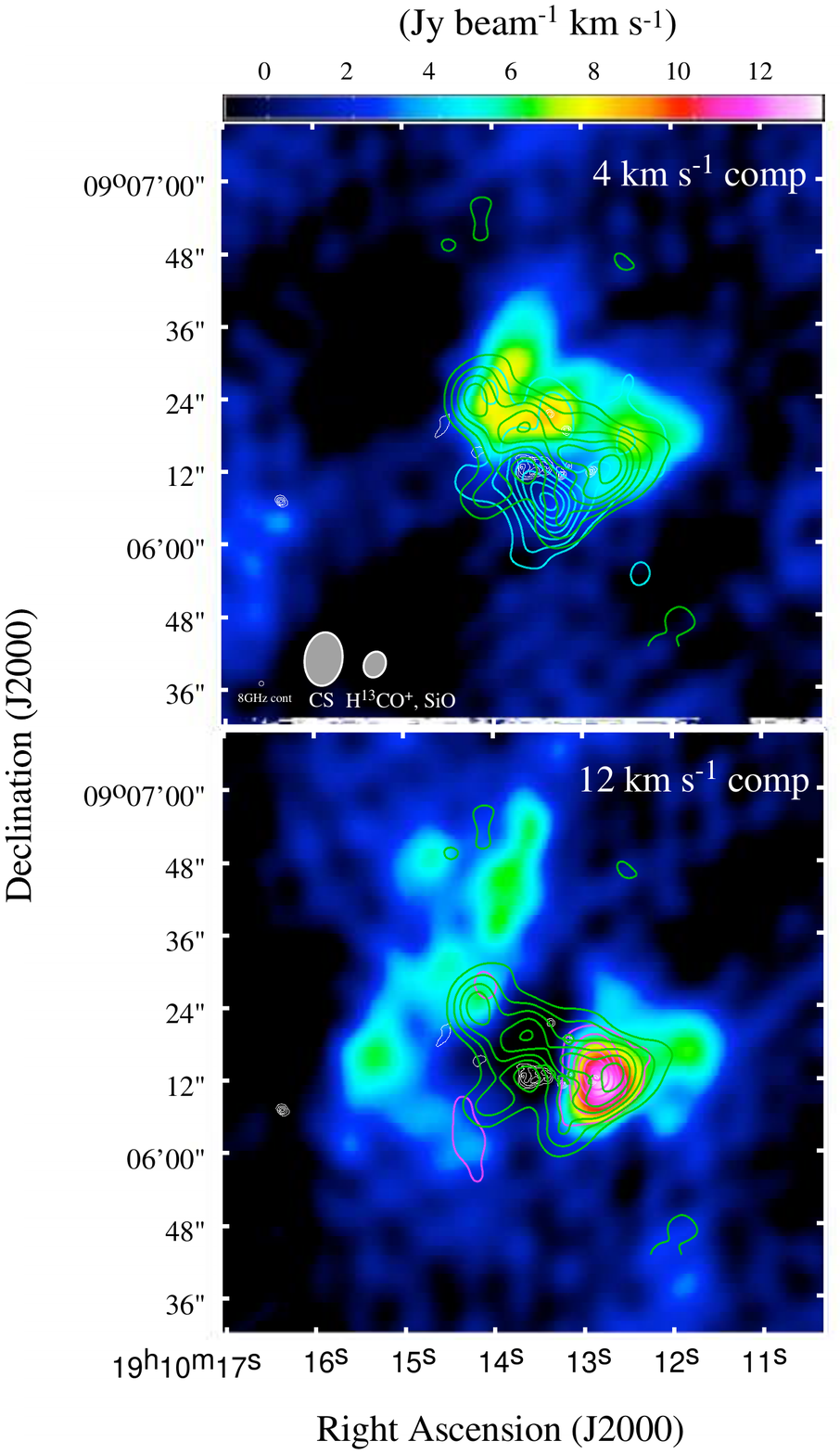}
\caption{Comparison of CS, H$^{13}$CO$^+$ and SiO distributions. 
{\it Upper panel:} color map is for the CS 4~km\,s$^{-1}$ component (integrated from $V_{\rm LSR}=-$5.0~km\,s$^{-1}$ to 7.4~km\,s$^{-1}$), blue contours are for H$^{13}$CO$^+$ 4~km\,s$^{-1}$ component (integrated from $-$4.4~km\,s$^{-1}$ to 7.5~km\,s$^{-1}$), and green contours are for the SiO 8~km\,s$^{-1}$ emission integrated from 5.0~km\,s$^{-1}$ to 10.4~km\,s$^{-1}$. 
The beam sizes are shown at the bottom left corner.
{\it Lower panel:} color map is for the CS 12~km\,s$^{-1}$ component (integrated from 8.3~km\,s$^{-1}$ to 19.8~km\,s$^{-1}$), red contours are for H$^{13}$CO$^+$ 12~km\,s$^{-1}$ component (integrated from 8.6~km\,s$^{-1}$ to 20.5~km\,s$^{-1}$), and green contours are for the SiO 8~km\,s$^{-1}$ emission.
Contours are drawn at the 20\%, 30\%, 40\%, 50\%, 60\%, 70\%, 80\%, and 90\% levels.
For both panels, the 8.3~GHz continuum image \citep[][]{DePree1997} is superposed in white contours.
\label{Fig15}}
\end{figure}

We summarize in Figure~\ref{Fig15} the positional relation between the UCHII, CS, H$^{13}$CO$^+$ and SiO emission features.
Overall, the SiO emission fills, or coincides with, the UCHII ring and is well delineated by the ring, and is roughly surrounded by the CS and H$^{13}$CO$^+$ emission clumps.
With respect to the CS emission, the SiO emission is located slightly to the south of its 4~km\,s$^{-1}$ component and between the northeast C-shaped feature (CS-NE12, CS-NE12A, CS-NE12B) and the southwest feature (CS-SW12, CS-NW12) of the 12~km\,s$^{-1}$ component.
With respect to the H$^{13}$CO$^+$ emission, the SiO emission  is located to the north of its 4~km\,s$^{-1}$ component and to the northeast of its 12~km\,s$^{-1}$ component.

\begin{figure}[htbp]
\includegraphics*[bb= 280 130 600 500, scale=0.75]{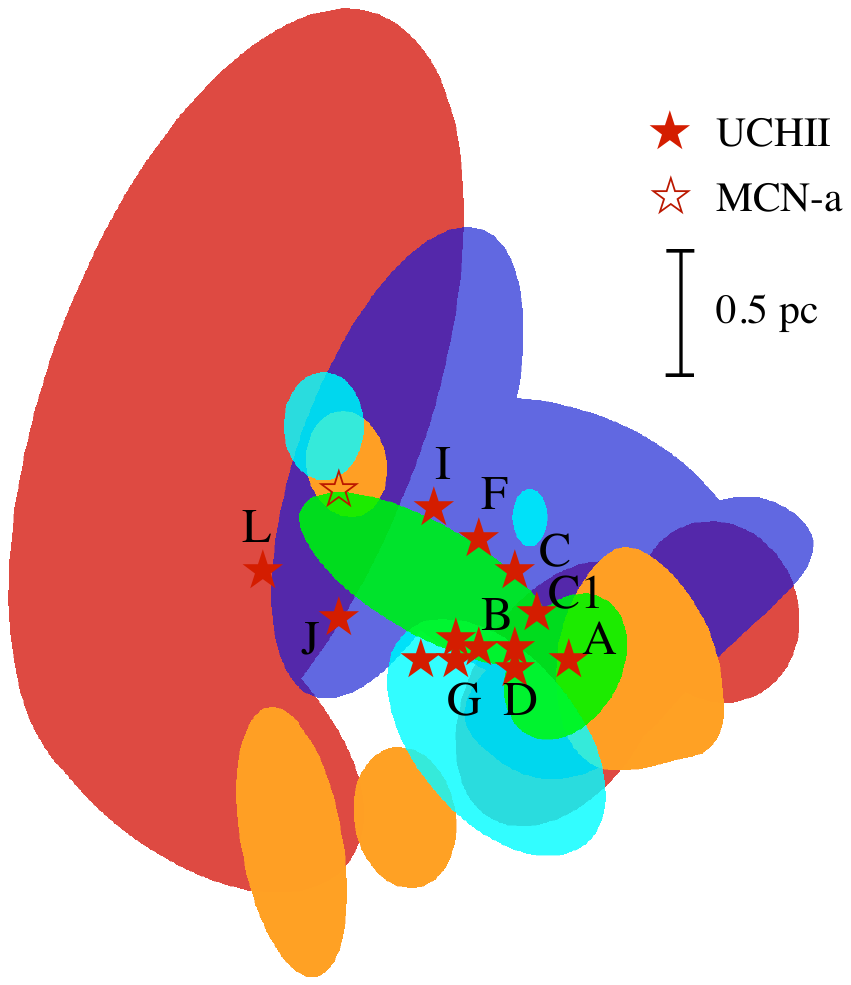}
\caption{
Schematic representation of gas distribution toward W49N. 
{\it Blue:} CS gas at the velocity of $\sim$4~km\,s$^{-1}$.
{\it Light blue:} H$^{13}$CO$^+$ gas at $\sim$4~km\,s$^{-1}$.
{\it Green:} SiO gas at $\sim$8~km\,s$^{-1}$.
{\it Red:} CS gas at $\sim$12~km\,s$^{-1}$.
{\it Orange:} H$^{13}$CO$^+$ gas at $\sim$12~km\,s$^{-1}$.
Filled asterisks denote UCHII regions and the open asterisk indicates MCN-a \citep{Wilner2001}.
\label{Fig16}}
\end{figure}

In summary, the spatial distribution of the 4~km\,s$^{-1}$ component of CS and H$^{13}$CO$^+$ agrees well with those of the UCHII ring and SiO gas, and the 12~km\,s$^{-1}$ component of CS and H$^{13}$CO$^+$ surrounds the UCHII ring and SiO gas.
We also note that, as seen in the PV diagrams (Figures~\ref{Fig07} and \ref{Fig08}), the characteristic radial velocity of SiO gas is between the 4~km\,s$^{-1}$ and 12~km\,s$^{-1}$ components of CS and H$^{13}$CO$^+$ gasses.
We illustrated in Figure~\ref{Fig16} a schematic diagram to visualize the configuration.

These characteristics are consistent with the cloud-cloud collision view of W49N.
A smaller cloud of $\sim$10$^4$~M$_\odot$ at a radial velocity of $V_{\rm LSR}\sim$4~km\,s$^{-1}$ has just collided, along the line of sight, with a larger one with a similar mass at $V_{\rm LSR}\sim$12~km\,s$^{-1}$, making a hole near the center of the original $\sim$12~km\,s$^{-1}$ cloud (see the top panel of Figure~\ref{Fig09}).
The original 4~km\,s$^{-1}$ cloud is seen still located in the hole, where the ring of UCHII regions and SiO gas exist.

Cases of cloud-cloud collision to trigger massive star formation have extensively been studied by \citet{Fukui2018} recently.
Their model calculations and compiled data show two observable features of cloud-cloud collision possibly seen in the early phases of massive star formation: complementary distribution of two colliding clouds, and bridge features in PV diagrams connecting the two velocity components of the colliding clouds.
In our data, the 4~km\,s$^{-1}$ and 12~km\,s$^{-1}$ components of CS show complementary distribution that the former component sits between, or inside, the two fragments of the latter.
This configuration also implies that the collision takes place almost along the line of sight.
The positional relation is particularly obvious in the PV diagram of CS along the major axis of the ring (see Figure~\ref{Fig07}).
In addition, the PV diagram shows faint CS emission between the two velocity components as indicated by arrows in Figure~\ref{Fig07}.
Such faint CS emission connecting the two velocity components may correspond to the bridge features pointed out by \citet{Fukui2018}.

Regarding the bridge feature, we should rather attribute the entire SiO emission to a bridge feature in our case.
The SiO emission is observed only toward hot ($>$\,30~K), dense regions heated by shocks \citep[e.g.,][]{Ziurys1989} associated with star forming activities, such as molecular outflows.
If a massive protostar forms in the shocked gas produced by cloud-cloud collision, it should naturally be surrounded by the gas with the intermediate velocity.
Successive activities, such as outflows, of the forming star produces additional shocks, from which SiO line is emitted, with a velocity near that of the original shocked gas , i.e., between the two colliding cloud velocities.
In the case of W49N, the SiO emission has a characteristic velocity of 6--8~km\,s$^{-1}$, which is between the two colliding cloud velocities of 4~km\,s$^{-1}$ and 12~km\,s$^{-1}$, consistent with the properties of bridge features.
In other words, the bridge feature is enhanced in the SiO emission.

Optically thick CS ($J=2-1$) lines show absorption (inverse P Cygni profiles) at 8~km\,s$^{-1}\,\la\,V_{\rm LSR}\,\la\,$20~km\,s$^{-1}$ against the strong continuum source~G \citep{Dickel1999}.
The redshifted absorption requires low temperature, low-density gas in front of source~G \citep{Williams2004}, suggesting that diffuse 12~km\,s$^{-1}$ gas is located in front of the UCHII ring, while the 4~km\,s$^{-1}$ gas is behind it.
This implies that the UCHII ring formed on the far side of the original 12~km\,s$^{-1}$ gas and on the near side of the original 4~km\,s$^{-1}$ gas.
Thus the central part of the original 12~km\,s$^{-1}$ cloud would be  just being penetrated, from its far side to near side, by the 4~km\,s$^{-1}$ cloud, and the star formation may be occurring at the interface, traced by the SiO emission, between the two clouds.

\citet{Miyawaki1986} suggested from the ages of the UCHII regions that the burst of star formation in W49N began in 10$^4$--10$^5$~yr ago. 
Dividing the spatial scale of the region ($\sim$1\,pc) by the relative velocity (8~km\,s$^{-1}$) of the two colliding clouds, we estimate, assuming a head-on collision along the line of sight, that the onset of collision took place $\sim$10$^5$\,yr ago, which agrees with the above starting time of the star burst in this region. 
The estimated free fall time scale $\sim3\times10^4$~yr of the clumps is shorter than, or at least comparable to, this duration of collision, and a few dozens of UCHII regions have been produced by the present time, with additional dozens being poised to form in the next 10$^5$\,yr.

A burst of star formation or a mini starburst possibly induced by cloud-cloud collision similar to W49N is also proposed for Sgr~B2 \citep[e.g.,][]{Hasegawa1994, Sato2000}, where a hole with shocked gas produced by collision was also observed, and W51A \citep[e.g.,][]{Okumura2001} in addition to those presented in \citet{Fukui2018}.
Such collisions between clouds naturally tend to form top-heavy initial mass functions as shown in recent calculations \citep[e.g.,][]{Fukui2020}.

\subsubsection{Global collapse view}\label{Global}

Let us discuss the global collapse model of W49N \citep{Welch1987} and its relation to our  cloud-cloud collision view based on the results that we so far obtained. 
\citet{Dickel1999} argued that the inverse P Cygni absorption of the CS ($J=2-1$) line profiles gave support to the global collapse model: the gas in front of W49N UCHII regions is infalling toward them.
\citet{Williams2004}, however, found that any of the three models they considered, namely global collapse of a very large (5 pc radius) cloud, localized collapse from smaller (1 pc) clouds around individual HII regions, and multiple, static clouds, can reproduce the CS ($J=2-1$) line profiles reasonably well provided that the component clouds have a core-envelope structure with a temperature gradient.
They presumed the core-envelope structure in order to reproduce the CS absorption by the low density, low temperature envelope against the continuum sources, while, at the same time, keeping the higher rotational levels of CS sufficiently populated.
Thus the presence of redshifted foreground gas does not provide  firm evidence for the global collapse model at a scale of 5 pc ($\sim$1.5$'\sim\,$FOV of CS maps of this paper).

In addition, the current high resolution CS ($J=1-0$) map does not confirm any kinematical trend of infall over its FOV: they show several clumps at $\sim$4~km\,s$^{-1}$ and $\sim$12~km\,s$^{-1}$ with no indication of ordered motion.
The radial velocities of recombination lines emitted from the UCHII regions do not show any ordered motion, such as rotation or infall of the UCHII ring, either \citep{DePree1997}.
We did not find any evidence to support the global collapse model of W49N at a scale of 5 pc.
This does not exclude localized collapse occurring at a scale of \la1\,pc in the clumps we identified in this paper as was discussed above.

Global collapse at a larger scale, e.g., \ga10 pc, could be possible. 
\citet{Madrid2013} presented CO ($J=1-0$) maps of the entire W49 giant molecular cloud (GMC) covering an area of $\sim35'\times35'$ (113 pc) obtained with the Purple Mountain Observatory 14 m telescope.
They found that the mass of the GMC is distributed in a hierarchical network of filaments, which are connected to the centrally condensed structure W49N at scales $<$10 pc, concluding that the W49A starburst most likely formed from global gravitational contraction with localized collapse in a ``hub-filament'' geometry.

The presence of molecular filaments converging to W49N may reflect a large scale gravitational infall at a scale of 100 pc.
At scales of $\la\,$20 pc, however, \citet{Miyawaki2009} did not find any motion indicative of systematic infall among the 14 $^{13}$CO features they identified, which may mean that molecular gas in the 100~pc scale infalling flow has fragmented into smaller clouds at this scale, followed by their rather random, turbulent motion.
Such clouds have a significant chance to collide with each other, as discussed by \citet{Miyawaki2009}, triggering the starburst in W49N.

\section{Conclusions}\label{Conclusions}

We have presented observations of CS ($J=1-0$), H$^{13}$CO$^+$ ($J=1-0$), and SiO ($v=0: J=1-0$) lines, together with the 49~GHz and 86~GHz continuum emissions, toward W49N carried out with Nobeyama Millimeter Array.
We identified 11 CS, 8 H$^{13}$CO$^+$, and 6 SiO clumps with radii of 0.1--0.5~pc by eye inspection on the velocity channel maps.
We calculated the column densities, masses, molecular hydrogen densities, and free fall time scales of the clumps.
We also derived the mass accretion rates of the clumps assuming that they are undergoing free fall.
We then discussed the cloud-cloud collision view in comparison with the global-collapse model for the burst of massive star formation in W49N.

Our main results are summarized as follows.
\begin{enumerate}

\item 
The CS and H$^{13}$CO$^+$ clumps are mainly divided into two velocity components, one at 4~km\,s$^{-1}$ and the other at 12~km\,s$^{-1}$.
The CS maps show spatially complementary distribution between the two velocity components: the 4~km\,s$^{-1}$ component clumps, coinciding with the UCHII regions, are located between, or inside, the 12~km\,s$^{-1}$ component clumps.
The position velocity diagrams of CS show faint emission features bridging the two velocity components.

\item 
The SiO clumps are distributed toward, or inside, the UCHII ring, where the 4\,km\,s$^{-1}$ component clumps of CS and H$^{13}$CO$^+$ also exist.
The SiO clump velocities are between the two velocity components of CS and H$^{13}$CO$^+$ clumps, corresponding to those of the bridge features.

\item 
The CS clump masses vary from $8.9\times10^2$~M$_{\odot}$ to $2.8\times10^4$~M$_{\odot}$ with the mean of $9.4\times10^3$ M$_{\odot}$.
The H$^{13}$CO$^+$ clump masses vary from $4.4\times10^2$~M$_{\odot}$ to $3.9\times10^4$~M$_{\odot}$ with the mean of $8.8\times10^3$ M$_{\odot}$.
The SiO clump masses vary from $4.7\times10^3$~M$_{\odot}$ to $4.9\times10^4$~M$_{\odot}$ with the mean of $2.2\times10^4$ M$_{\odot}$.
These values have a typical error of a factor of several times.
The derived masses of the clumps suggest that they have the characteristics of massive star forming cloud fragments.

\item
The total masses derived from CS, H$^{13}$CO$^+$, and SiO clumps are 1.0$\times$10$^5$~M$_{\odot}$, 7.0$\times$10$^4$~M$_{\odot}$, and 1.3$\times$10$^5$~M$_{\odot}$, respectively, which can be compared with the corresponding virial masses of 7.1$\times$10$^4$~M$_{\odot}$, 1.3$\times$10$^5$~M$_{\odot}$, and 8.8$\times$10$^4$~M$_{\odot}$, respectively.
The two types of mass agree well with each other.

 \item
The average densities of the clumps are $9.0\times 10^5\,$cm$^{-3}$, $1.4\times 10^6\,$cm$^{-3}$, and $7.6\times 10^6\,$cm$^{-3}$ for the CS, H$^{13}$CO$^+$ and SiO clumps, respectively.
The density derived from SiO seems significantly higher than those derived from CS and  H$^{13}$CO$^+$, which may be due to the SiO emission produced in high density shocked regions.

\item
The free fall time scale of the clumps is estimated to be $\sim3\times10^{4}$~yr with a typical error of a factor of three.
This gives a nominal accretion rate, i.e., the clump mass divided by the free fall time, of 10$^{-2}$--3~M$_\odot$yr$^{-1}$, which leads to the actual accretion rate of 3$\times$10$^{-3}$--1~M$_\odot$yr$^{-1}$ onto a stellar core.
The observed clumps are, if they are undergoing free fall, capable of producing massive stars in the next 10$^5$~yr.

\item 
We propose a view that a 4~km\,s$^{-1}$ cloud collided with a larger 12~km\,s$^{-1}$ cloud leaving a hole at the center of the original 12~km\,s$^{-1}$ cloud.
The collision produced shocked, intermediate velocity gas, seen as bridge features in the PV diagrams, where dozens of unstable clumps of 10$^4$~M$_\odot$ formed, and triggered a burst of massive star formation in W49N.

\end{enumerate}

\begin{ack}\label{Acknowledgement}
We are deeply indebted to the NRO staff for their operation of telescopes and continuous efforts to improve the performance of the instruments.
In particular, the authors wish to thank late Professor K.-I. Morita for his indispensable advice about preparing for and carrying out observations and reducing millimeter-wave interferometer data.
We are grateful to the anonymous referee who provided us with useful feedback to improve the paper.

\end{ack}

\begin{appendix}\label{Appendix}

\section{Comparison of spectral line profiles and the two velocity components}
\label{line profiles}

\begin{figure*}[htbp]
\includegraphics*[bb= 150 100 900 500, scale=0.75]{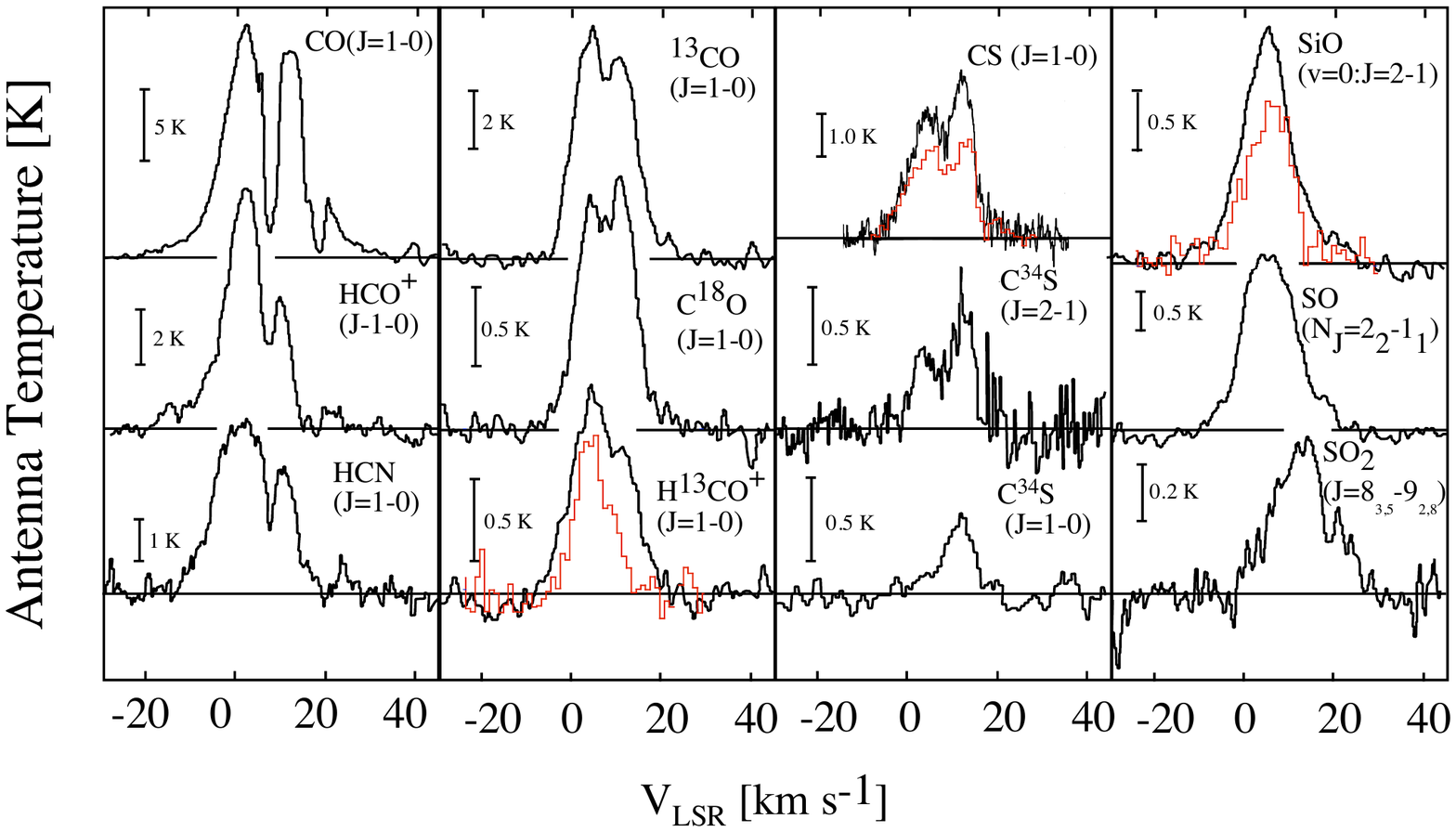}
\caption{
{\it Black lines:} Spectral line profiles of CO ($J=1-0$), $^{13}$CO ($J=1-0$),  C$^{18}$O ($J=1-0$), HCN ($J=1-0$), HCO$^+$ ($J=1-0$),  H$^{13}$CO$^+$\ ($J=1-0$), CS ($J=1-0$), C$^{34}S$ ($J=1-0$), 
C$^{34}$S($J=2-1$), SiO ($v=0:~J=2-1$), SO ({\it N$_J$}=2$_2$--1$_1$), and SO$_2$ ({\it J}=8$_{3,5}$--9$_{2,8}$) obtained with the 45-m telescope (see Appendix~1). 
{\it Red lines:} Spectral line profiles of  CS ({\it J}=1-0), H$^{13}$CO$^+$\ ($J=1-0$), and SiO (v=0; $J$=2--1) obtained with NMA averaged over the corresponding 45 m telescope beam sizes. Their specific intensities (brightness temperatures) were converted to antenna temperatures by multiplying the main beam efficiencies of 0.75 and 0.48 for the 49 GHz and 86 GHz lines, respectively, of the 45 m telescope: 
\label{Fig17}}
\end{figure*}

Figure~\ref{Fig17} compares the interferometric spectra (shown in red) of CS ($J=1-0$), SiO ($v=0:~J=2-1$) and H$^{13}$CO$^+$\ ($J=1-0$) with those of the same and other lines (shown in black) obtained with the Nobeyama 45 m telescope (see Appendix~\ref{45 m observations}). 
The interferometric spectra are averaged over the corresponding beam sizes of the 45 m telescope.
The intensity scale of 0.2 Jy\,beam$^{-1}$ corresponds to 1.0~K  for the CS ($J=1-0$) emission and 3.5~K for the  H$^{13}$CO$^+$ and SiO emission, respectively.

From the compilation in  Figure~\ref{Fig17}, we notice that there are generally three types of spectral line profiles: those showing double peaks with a deep dip at $\sim$7~km\,s$^{-1}$, those with a shallow dip near the line center, and those with a single peak.   
We call them as ``deep dip lines'', ``shallow dip lines,'' and ``single peak lines,'' respectively, in this paper.

CO ($J=1-0$), HCO$^+$ ($J=1-0$), and HCN ($J=1-0$) are classified into the deep dip lines.
They are characterized by a deep dip at $\sim$7~km\,s$^{-1}$ probably caused by foreground absorbing gas, consistent with the large optical depths of these emission lines.

$^{13}$CO ($J=1-0$), C$^{18}$O ($J=1-0$), H$^{13}$CO$^+$\ ($J=1-0$), and C$^{34}S$ ($J=2-1$) are the shallow dip lines.  
The $^{13}$CO and C$^{18}$O line profiles are similar in shape: they have two peaks at $V_{\rm LSR}$$\sim$4 and 12~km\,s$^{-1}$\ with roughly the same intensity.  
A shallow dip is seen between the two peaks at $\sim$7~km\,s$^{-1}$.   
The $^{13}$CO intensity is stronger than the C$^{18}$O\ by a factor of 5, close to the ordinary abundance ratio of the two isotopes, suggesting that both these lines are optically thin.   
The single dish H$^{13}$CO$^+$ line profile also has two peaks at $V_{\rm LSR}$$\sim$4 and 12~km\,s$^{-1}$, but with a shallow dip at $\sim$9~km\,s$^{-1}$.  
The blueshifted peak is significantly stronger than the redshifted peak in this case. 
The NMA spectrum of H$^{13}$CO$^+$ averaged over the 45 m telescope beam shows a single peak at $\sim$4~km\,s$^{-1}$ and has relatively weak emission at $\sim$12~km\,s$^{-1}$, suggesting that the emission region is more extended at $\sim$12~km\,s$^{-1}$ than at $\sim$4~km\,s$^{-1}$.

The line profiles of C$^{34}$S ($J=2-1$) and C$^{34}$S ($J=1-0$) may also be classified into the shallow dip lines.
The C$^{34}$S ($J=2-1$) line has double peaks at $V_{\rm LSR}$$\sim$4 and 12~km\,s$^{-1}$ with a shallow dip at $\sim$9~km\,s$^{-1}$. 
Its redshifted peak is stronger than the blueshifted peak, on the contrary to the single dish H$^{13}$CO$^+$ line profile.
The C$^{34}$S ($J=1-0$) line shows a single peak at $V_{\rm LSR}$$\sim$12~km\,s$^{-1}$\ with an indication of a secondary feature at $\sim$4~km\,s$^{-1}$. 
By analogy with the C$^{34}$S ($J=2-1$) line profile, we may naturally think that C$^{34}$S ($J=1-0$) also has a double peak characteristic, but its blueshifted peak would be much weaker than the red one and is not detected.

As is exemplified from the ratio of $^{13}$CO to C$^{18}$O intensity, the shallow dip lines are optically thin.
This is consistent with the view that there are two velocity components in W49N corresponding to the two peaks.
For ``shallow dip lines,'' the overall shape of a line profile and the velocity of a dip, if any, are then determined by the relative strength of the two velocity components.
The HNC ($J=1-0$) line \citet{Nyman1989} is also classified into this type. 

The case for the main isotopologue of CS needs some caution, because the emission may be optically thick.
Both the single dish and NMA line profiles of CS ($J=1-0$) have a relatively shallow dip at $V_{\rm LSR}\,\sim$8~km\,s$^{-1}$ similar to C$^{34}$S ($J=2-1$).
The CS ($J=2-1$) line profile toward W49N \citep{Nyman1984} are also similar to those of $^{13}$CO to C$^{18}$O.
In addition, the relative strength of CS and C$^{34}$S line profiles at $V_{\rm LSR}=$4~km\,s$^{-1}$ and 12~km\,s$^{-1}$ depends not only on positions, but on the rotational transitions observed \citep{Serabyn1993}, suggesting that there are two velocity components at $\sim$4~km\,s$^{-1}$ and $\sim$12~km\,s$^{-1}$ toward W49N with different physical conditions and spatial extent.
The CS ($J=1-0$) line should thus be categorized as a shallow dip line.

The SiO ($v=0:~J=2-1$) and SO ({\it N$_J$}=2$_2$--1$_1$) lines have a single peak at $V_{\rm LSR}$$\sim$6~km\,s$^{-1}$ and are classified as the single peak lines.
Thermal emissions of SiO and SO have so far been detected from highly excited regions such as the Orion hot core or molecular gas interacting with outflows \citep[cf.][]{Downes1982, Wright1983}.  
There is an increasing number of evidence that such molecules are evaporated from dust grains when grains are exposed to shocks. 
It is interesting that these shock indicators show quite different characteristics in their line profiles compared with the deep dip and shallow dip lines: their peak velocities are $\sim$6~km~s$^{-1}$, which is between the two peaks of the lines with dips. 

Although the SO$_2$ ({\it J}=8$_{3,5}$ -- 9$_{2,8}$) line profile also shows a single peak, its velocity is $\sim$13~km\,s$^{-1}$, similar to the 12~km\,s$^{-1}$ component.
In addition, \citet{Serabyn1993} noted that their SO$_2$ ({\it J}=18$_{1,17}$ -- 18$_{0,17}$) spectra show contribution from the same two velocity components as CS and C$^{34}$S.
The SO$_2$ ({\it J}=8$_{3,5}$ -- 9$_{2,8}$) line profile can thus be regarded as a case when the 4~km\,s$^{-1}$ component is much weaker than the 12~km\,s$^{-1}$ component.
We accordingly classify the SO$_2$ emission profile into the shallow dip lines.

\section{The 45 m telescope observations of W49N}
\label{45 m observations}

Observations of the spectra shown in Figure~\ref{Fig17} were made in various occasions between 1986 and 1989 using the 45 m telescope of NRO.   
The telescope beam size (HPBW) was $\sim$34\arcsec, $\sim20$\arcsec, $\sim$17\arcsec at 49 GHz, 86 GHz, and 110 GHz, respectively, with the typical pointing accuracy of 5\arcsec. 
The reference center for all the observations was taken to be (${\alpha(1950)}$, ${\delta(1950)}$) = (19$^{\rm h}$07$^{\rm m}$49\fs8, 9\degree01$'$15\farcs5), which is the position of  source~G \citep{Dreher1984}. 
The data were taken with acousto-optical spectrometers with a high-resolution (AOS-H) and a wide-band (AOS-W) capabilities.
AOS-H covers a 40 MHz bandwidth with a resolution of 37 kHz, corresponding to the velocity resolutions of 0.23~km\,s$^{-1}$ and 0.10--0.13~km\,s$^{-1}$ at 48~GHz and 86--110~GHz, respectively.
AOS-W covers a bandwidth of 250 MHz with a resolution of 250 kHz, corresponding to the velocity resolutions of 1.56~km\,s$^{-1}$ and 0.68--0.87~km\,s$^{-1}$ at 49~GHz and 86--110~GHz, respectively.
Some of the data such as CS ($J=1-0$) and C$^{34}$S ($J=1-0$) lines are presented and discussed in \citet{Miyawaki1986}.

\end{appendix}

\label{References}

\vspace{20mm}

\end{document}